\documentclass[12pt,english]{article}
\pdfoutput=1
\usepackage{jheppub,empheq}
\usepackage{simplewick}
\usepackage{verbatim}
\usepackage{subfigure}
\usepackage{graphics, color}
\usepackage{graphicx}
\usepackage{mathabx}
\usepackage{amsmath,mathtools}
\usepackage{amssymb}
\usepackage{amsthm}
\usepackage{appendix}
\usepackage{amsfonts}
\usepackage{lipsum}   
\usepackage[usenames,dvipsnames]{xcolor}
\usepackage[normalem]{ulem}
\usepackage{braket}

\definecolor{darkgreen}{rgb}{0,0.5,0}

\newcommand{\be}{\begin{equation}}
\newcommand{\ee}{\end{equation}}

\newcommand{\zb}{\bar{z}}
\newcommand{\dDisc}{\text{dDisc}}

\newmuskip\pFqmuskip
\newcommand*\pFq[6][8]{%
  \begingroup 
  \pFqmuskip=#1mu\relax
  \mathcode`\,=\string"8000
  \begingroup\lccode`\~=`\,
  \lowercase{\endgroup\let~}\pFqcomma
  {}_{#2}\tilde{F}_{#3}{\left[\genfrac..{0pt}{}{#4}{#5};#6\right]}%
  \endgroup
}
\newcommand{\pFqcomma}{\mskip\pFqmuskip}

\graphicspath{ {./images/} }

\def\({\left(}
\def\){\right)}

\begin{document}

\title{Lorentzian inversion and anomalous dimensions in Mellin space}
\author{Milind Shyani}
\affiliation{Stanford Institute for Theoretical Physics, Department of Physics, Stanford University, Stanford, CA 94305, USA}
\emailAdd{shyani@stanford.edu}
\abstract{In this note, we derive a Mellin space version of the Lorentzian inversion formula for CFTs by explicitly integrating over the cross-ratios in $d=2$ and $d=4$ spacetime dimensions. We use the simplicity of the Mellin representation of Witten diagrams and the double discontinuity to find the OPE coefficients and anomalous dimensions of double-trace primaries in large $N$ CFTs to order $\frac{1}{N^4}$. We find that our results match analytically at order $\frac{1}{N^2}$, and numerically at order $\frac{1}{N^4}$ with existing literature.
\\
\\ 
\begin{center}
{\it Dedicated to the memory of Joe Polchinski.}
\end{center}
}
\maketitle
\addtocontents{toc}{\protect\enlargethispage{2\baselineskip}}

\section{Introduction}

One of the recent breakthroughs in the bootstrap program \cite{Poland:2018epd,Simmons-Duffin:2016gjk} for conformal field theories in spacetime dimensions $d>2$ has been the discovery of the Lorentzian inversion formula by Caron-Huot \cite{Caron-Huot:2017vep,Simmons-Duffin:2017nub}. The inversion formula provides an analytic formula for the OPE coefficients in terms of the double-discontinuity (dDisc) of the four-point function $\mathcal G(z,\zb)$. In theories with a small parameter, like the Wilson-Fisher theory or a large $N$ CFT, dDisc$\left[\mathcal G(z,\zb)\right]$ is easier to calculate than the four-point function $\mathcal G(z,\zb)$ itself. This feature in addition to the analyticity of the inversion formula provides a new powerful tool in the bootstrap program. Since its recent discovery, the inversion formula has already given several new insights about conformal field theories \cite{Alday:2017vkk,Alday:2017zzv,Caron-Huot:2018kta,Kravchuk:2018htv,Iliesiu:2018fao,Mukhametzhanov:2018zja}. 

Another important progress has been the recent approach to conformal bootstrap in Mellin space \cite{Gopakumar:2016cpb,Gopakumar:2018xqi,Polyakov:1974gs}. This new revival, along with the well known simplicity of the Mellin representation of Witten diagrams  \cite{Penedones:2010ue,Costa:2012cb,Fitzpatrick:2011ia,Fitzpatrick:2011hu,Fitzpatrick:2011dm,Paulos:2011ie,Cardona:2017tsw,Rastelli:2016nze,Yuan:2018qva} motivated us to look for an inversion formula in Mellin space. One of the first simplifications in Mellin space is the apparent ease of taking the double-discontinuity of $\mathcal{G}(z,\zb)$. The $\dDisc_t$ produces zeros that exactly cancel the double-trace $t$-channel poles of the gamma functions present in the Mellin measure. This feature has also been noticed in the work by Cardona \cite{Cardona:2018nnk}, where the inversion formula has been considered in the collinear approximation to obtain interesting results. 

In this note however, we do not work in any limit and explicitly integrate out the cross ratios to obtain what we call the Mellin inversion formula in $d=2$ (equation \ref{Mellin2d}) and in $d=4$ (equation \ref{Mellin4d}). Exchanging the order of integration over the cross ratios $(z,\zb)$ and the Mellin variables $(s,t)$ requires some care, and indeed we find that the naive inversion kernel in Mellin space blows up for several values of $s$ and $t$. This problem is surmountable, precisely due to the analyticity of the inversion formula. We show in section \ref{melcon}, how a beautiful identity involving the hypergeometric function ${}_3F_2$ at unit argument from \cite{Buhring:1987:BUA:36949.36951}\footnote{We thank Raghu Mahajan for pointing out this paper to us.} helps in analytically continuing our results to all values of $s$ and $t$.

The problem of solving the bootstrap equations for large $N$ theories isn't new. Starting with the seminal work of \cite{Heemskerk:2009pn}, there has been significant progress in understanding the structure of large $N$ theories \cite{Heemskerk:2010ty,Fitzpatrick:2012cg,ElShowk:2011ag,Alday:2014tsa,Hijano:2015zsa,Alday:2017gde} via the conformal bootstrap. In recent years, several new methods have been found (some of which use the inversion formula) to obtain the OPE coefficients and anomalous dimensions in large $N$ theories \cite{Aharony:2016dwx,Liu:2018jhs,Cardona:2018dov,Cardona:2018qrt,Sleight:2018epi,Sleight:2018ryu,Zhou:2018sfz,Ghosh:2018bgd}. The aim of this work is to provide a new perspective of obtaining such results using the power of the inversion formula and the simplicity of the Mellin space representation of AdS correlators. The Mellin inversion formulas (\ref{Mellin2d}) and (\ref{Mellin4d}), provide the OPE coefficients in terms of the Mellin amplitude. We use the Mellin amplitudes of Witten diagrams to calculate the OPE coefficients and anomalous dimensions of leading-twist double-trace primaries. We find that our results match perfectly with those in the literature \cite{Fitzpatrick:2012yx,Komargodski:2012ek,Liu:2018jhs,Aharony:2016dwx}. 

We start with the contact Witten diagram in a scalar bulk theory with quartic vertices in section \ref{seccon}. The Mellin amplitude is just a constant and we find that the Mellin inversion formula gives a vanishing result. This seems obviously wrong, but we argue why this isn't unexpected due to the invalidity of the inversion formula for low enough spins. We discuss the region of convergence of the Mellin inversion formula for Witten diagrams in section \ref{melcon} and show that the contact Witten diagram lies outside this region. 

We then study the exchange Witten diagram in the bulk scalar theory with cubic vertices in section \ref{secexc}. We find that the inversion formula has poles corresponding to the double-trace primaries 
\begin{align}
    \Delta = \Delta_1 + \Delta_2 + J + 2n,  \quad \Delta = \Delta_3 + \Delta_4 + J + 2n, \qquad n=0,1,2, \ldots \, . \label{doub}
\end{align}
The residue of the inversion formula at these poles provide the OPE coefficients and anomalous dimensions of the double-trace primaries at $O(1/N^2)$. This result has also been found recently by Liu et al. \cite{Liu:2018jhs} and we find that our results match exactly.

Finally, we study the bubble diagram in the the bulk scalar theory with quartic vertices in section \ref{secbub}. We again find that the inversion formula has poles at locations given in (\ref{doub}) corresponding to the double-trace primaries. The residue at these poles provide the OPE coefficients and anomalous dimensions of the double-trace primaries at $O(1/N^4)$. As far as we know, results at this order have been only calculated recently, starting with the work of Aharony et al. \cite{Aharony:2016dwx} for special values of the operator scaling dimensions. We find that our results match numerically and extend their results to arbitrary values of the scaling dimensions. We also find that our results match analytically for large $J$.  

Another interesting reason to study the inversion formula in Mellin space is the ease of taking the flat space limit of AdS. It has now been well established following the seminal work of Penedones \cite{Penedones:2010ue,Fitzpatrick:2011hu,Paulos:2016fap}, that in the large $(s,t)$ limit the Mellin amplitude $A(s,t)$ can be expressed as a flat space scattering amplitude $\mathcal T(s,t)$. Understanding this limit might provide a helpful prescription of connecting the CFT inversion formula with the flat space QFT Froissart-Gribov formula.  The flat space limit has also been understood in position space \cite{Gary:2009ae,Maldacena:2015iua}, but the simplicity of Mellin space correlators even with stringy corrections \cite{Alday:2018kkw,Alday:2018pdi,Rastelli:2016nze}, might make the flat space limit via Mellin space more transparent. We end this note in section \ref{sec4} with discussions about the flat space limit of our formula and future directions. The appendices include details about the normalisation of exchange diagrams, and calculations about analytically matching our results at order $\frac{1}{N^2}$ with those in the literature.


\section{Lorentzian inversion in Mellin space} \label{sec2}
Consider the four-point function of external scalars $\phi_i$ with conformal dimension $\Delta_i$,
\begin{align}
    G(x_i) = \braket{\phi_1(x_1)\phi_2(x_2)\phi_3(x_3)\phi_4(x_4)}. \nonumber
\end{align}
Using conformal symmetry we can rewrite this as
\begin{align}
     G(x_i) = \frac{1}{|x_{12}|^{\Delta_1 + \Delta_2} |x_{34}|^{\Delta_3 + \Delta_4}} \left( \frac{x^2_{14}}{x^2_{24}} \right)^a \left(  \frac{x^2_{14}}{x^2_{13}} \right)^b \mathcal G(u,v), \nonumber
\end{align}
where $u$ and $v$ are the conformal cross ratios,
\begin{align}
    u=z\zb= \frac{x^2_{12} x^2_{34}}{x^2_{13} x^2_{24}} , \quad v=(1-z)(1-\zb)=\frac{x^2_{14} x^2_{23}}{x^2_{13} x^2_{24}}, \label{cross}
\end{align}
and 
\begin{align}
    a  =\frac 1 2 \left(\Delta_2 - \Delta_1 \right), \quad b=\frac 1 2 \left( \Delta_3 - \Delta_4 \right). \nonumber
\end{align}
We will call $\mathcal G(u,v)$ the stripped four-point function, and it can be expanded in terms of conformal blocks,
\begin{align}
    \mathcal G(u,v) = 
    \sum_{\Delta,J} a_{\Delta,J} g_{\Delta,J}(u,v) \, .
    \label{sconf}
\end{align}
The closed form expressions of conformal blocks are only known in even spacetime dimensions. In this note we work in $d=2$ and $d=4$ spacetime dimensions. The blocks are given in terms of the hypergeometric functions,
\begin{align}
    g^{4d}_{\Delta,J}  &=  \frac{z \zb}{\zb - z}  \left( k_{\Delta-J-2}(z)  k_{\Delta+J}(\zb) -  k_{\Delta+J}(z)  k_{\Delta-J-2}(\zb) \right), \nonumber \\
    g^{2d}_{\Delta,J} &=  \frac{1}{1+\delta_{J,0}}\left(k_{\Delta-J}(z) k_{\Delta+J}(\zb) + k_{\Delta+J}(z)k_{\Delta-J}(\zb)\right), \nonumber \\
    k_{\beta}(z) &= z^{\beta/2} {}_2 F_1 \left( \frac \beta 2 + a, \frac \beta 2 +b,\beta;z\right).  
    \label{def} 
\end{align}
The conformal blocks are completely fixed by conformal symmetry while the OPE coefficients $a_{\Delta,J}$ contain the dynamical information of the theory. We will collectively refer to the OPE coefficients $a_{\Delta,J}$ and the operator spectrum $(\Delta,J)$ as the CFT data. The inversion formula presents an analytic function in $J$ that encodes the $s$-channel CFT data via its residues and poles,
\begin{align}
    c(\Delta,J) \sim \frac{a_{\tilde\Delta,J}}{\tilde\Delta-\Delta}, \label{cope}
\end{align}
where $\tilde\Delta$ is the dimension of a physical operator in the theory. It is given in terms of the $t$ and $u$-channel data,
\begin{align}
    c(\Delta,J) = c^t(\Delta,J) + (-1)^J c^u(\Delta,J), \label{t+u}
\end{align}
where
\begin{align}
    c^t(\Delta,J)= \frac{\kappa_{\Delta+J}}{4} 
    \int_0^1 dz \int_0^1 d\bar z \, \mu(z,\zb) \, g_{J+d-1,\Delta+1-d}(z,\bar z) 
    \, \mathrm{dDisc}_t[\mathcal G(z, \bar z)]. \label{cinv}
\end{align}
$c^u(\Delta,J)$ is given by the same formula but with the integration ranging from $-\infty$ to $0$, and the $\dDisc_u$ taken around $z=\infty$. The constant prefactor and the measure are given by,
\begin{align}
\kappa_{\Delta+J} & =  \frac{\Gamma\left(\frac{\Delta + J}{2} -a \right)\Gamma\left(\frac{\Delta + J}{2} +a \right) \Gamma\left(\frac{\Delta + J}{2} -b \right) \Gamma\left(\frac{\Delta + J}{2} +b \right)}{ 2\pi^2 \Gamma(\Delta + J -1) \Gamma(\Delta + J)}, \nonumber \\
\mu(z,\zb) & = \left |  \frac{z-\zb}{z \zb}   \right |^{d-2} \frac{\left((1-z)(1-\zb)\right)^{a+b}}{(z \zb)^2}\, .
\end{align}
The $\dDisc_t$ of the stripped four-point function is defined as,
\begin{align}
\mathrm{dDisc}_t[\mathcal G(z,\zb)] & \equiv \cos(\pi (a+b ))\mathcal G(z,\zb) - \frac{e^{i \pi (a+b)}}{2}\mathcal G\left(z,\zb \right)^\circlearrowleft - \frac{e^{-i \pi (a+b)}}{2}\mathcal G\left(z, \zb \right)^\circlearrowright ,\label{dDisc}
\end{align}
where the circle arrows represent going around $\zb =1$. Equation (\ref{cinv}) is the famous Lorentzian inversion formula in $d$ spacetime dimensions \cite{Caron-Huot:2017vep,Simmons-Duffin:2017nub}.

One of the essential reasons behind the utility of the Lorentzian inversion formula is that it only requires the dDisc[$\mathcal G(z,\zb)$] and not the entire four-point function $\mathcal G(z,\zb)$. It is well known that the double discontinuity $\dDisc_t[\mathcal G(z,\zb)]$ gets no contribution from operators that satisfy,
 \begin{align}
    \Delta = \Delta_2 + \Delta_3 + J + 2n, \qquad \Delta = \Delta_1 + \Delta_4 + J + 2n, \label{poleg}
\end{align}
a fact that we will also demonstrate in Mellin space. The important point to notice here is that these are precisely the operator dimensions of composite primary operators,
\begin{align}
    \left [ \phi_i \phi_j \right]_{n,J} \equiv \phi_i \nabla_{\mu_1} \nabla_{\mu_2} \ldots \nabla_{\mu_J} \left(\nabla^2\right)^n \phi_j + \ldots \, . \label{dop}
\end{align}
The dots correspond to similar terms that are added to make it a conformal primary. Such primaries are present in all theories that are perturbatively close to generalised free fields. Thus the contribution of these primaries can be safely ignored while evaluating the $\dDisc_t[\mathcal G(z,\zb)]$ in such theories. The Wilson-Fisher theory or large $N$ conformal field theories are the most prominent examples where the inversion formula can be put to immediate use. In a large $N$ theory, the operators (\ref{dop}) are called the double-trace primaries. 

The other important feature of the inversion formula is its analyticity in spin, $J$. This can be seen from (\ref{cinv}) by using the analytic properties of the conformal block and the fact that $z,\zb$ only range between 0 to 1. Similar arguments can be made for $c^u(\Delta,J)$. 

In the following section \ref{basic}, we review some basic facts about Mellin space. We then obtain the inversion formula in Mellin space by evaluating the $z,\zb$ integral in section \ref{meat}. We call (\ref{Mellin2d}) and (\ref{Mellin4d}) as the $d=2$ and $d=4$ Mellin inversion formula respectively. We end the section in \ref{melcon} with some important issues of convergence in Mellin space. Through out this note, we carry out the case of $d=2$ in detail. The calculations for $d=4$ are identical and we only state the final results.

\subsection{Basics of Mellin space} \label{basic}
The Mellin transform of the connected $n$-point conformally invariant correlator is given by,
\begin{align}
\braket{\phi_1(x_1)\phi_2(x_2) \ldots \phi_n(x_n)} = \frac{1}{(2\pi i)^{n(n-3)/2}} \int d\delta_{ij} A(\delta_{ij})\prod_{i<j}^{n} \Gamma(\delta_{ij})(x^2_{ij})^{-\delta_{ij}}, \label{master}
\end{align}
where the $\delta_{ij}$ satisfy the constraints $\sum_{i \neq j} \delta_{ij} = \Delta_j$. The integration contours run parallel to the imaginary axis and are placed on the real axis such that the poles from the gamma functions at $\delta_{ij}=-2k$, for non-negative integer $k$,  lie on one side. There are $\frac{n(n-1)}{2}-n$ free variables, which indeed are the number of independent conformal cross ratios of an $n$-point function. The stripped four-point function $\mathcal G(z,\zb)$ is a conformally invariant four-point function, and can also be expressed in Mellin space as,

{\small
\begin{align}
    \mathcal G(u,v) & = \int \frac{ds dt}{(2\pi i )^2} u^{\frac s 2} v^{\frac{t - \Delta_2 - \Delta_3}{2}}  \Gamma \left(\frac{\Delta_1 + \Delta_4 -t}{2} \right) \Gamma \left(\frac{\Delta_2 + \Delta_3 -t}{2} \right) \Upsilon(s,t)  A(s,t), \label{Mellin} \\
    \Upsilon(s,t) & \equiv  \Gamma \left( \frac{\Delta_1 + \Delta_2 -s}{2}\right) \Gamma\left(\frac{ \Delta_3 + \Delta_4 - s}{2} \right) \Gamma\left( \frac{s+t-\Delta_2 - \Delta_4 }{2} \right) \Gamma\left(\frac{s+t-\Delta_1 - \Delta_3 }{2}\right). \nonumber
\end{align}}

\noindent We use the conventions of Rastelli and Zhou \cite{Rastelli:2017udc} throughout this paper. The six gamma functions correspond to the double-trace poles in $s$, $t$ and $\hat u=\sum_i \Delta_i - s- t$. The reason why we have excluded the gamma functions corresponding to the $t$-channel double-trace poles in defining $\Upsilon(s,t)$ will become obvious in the next section.

A standard exercise to understand the structure of the Mellin amplitude $A(s,t)$ is to expand the four-point function in conformal blocks, (say) the s-channel. For e.g. in the limit $u \rightarrow 0, v \rightarrow 1$ for fixed $\frac{v-1}{\sqrt{u}}$, the left hand side of (\ref{Mellin}) becomes,
\begin{align}
    \mathcal G(u,v)  \approx \sum_{\Delta,J} a_{\Delta,J} u^{\frac{\Delta}{2}} C^{\frac{d}{2}-1}_J \left( \frac{v-1}{2\sqrt u } \right), \label{gegexp}
\end{align}
where $C_J$ are the Gegenbauer polynomials,
\begin{align}
    C^{\frac d 2 -1}_J(x)  = \sum_{m=0}^{\lfloor{J/2}\rfloor} c_m x^{J-2m}, \label{geg}
\end{align}
for some coefficients $c_m$ \cite{wolfram}. Comparing (\ref{Mellin}) with (\ref{gegexp}), we find that the Mellin amplitude should have poles at 
\begin{align}
s= \Delta - J+2m , \label{sdoub}
\end{align}
with the appropriate residues to reproduce the OPE coefficient and the Gegenbauer coefficients $c_m$. We refer the reader to the original work of Costa et al. \cite{Costa:2012cb} for more details where it is argued that\footnote{Our conventions (Rastelli and Zhou \cite{Rastelli:2017udc}) are different from those of Costa et al. \cite{Costa:2012cb}. $t_{\text{there}}=s_{\text{here}}, \, s_{\text{there}} = \Delta_2 + \Delta_3 - s_{here} - t_{here} $.},
\begin{align}
    A(s,t) \sim \frac{ a_{\Delta,J}\mathcal Q_{J,m}(\Delta_2  + \Delta_3 - s - t)}{s- \Delta+J-2m}, \qquad m=0,1,2, \ldots \, , \label{hahn}
\end{align}
where $\mathcal Q_{J,m}$ are related to what are known as the continuous Hahn polynomials \cite{Aharony:2016dwx,Gopakumar:2016cpb}. Equation (\ref{sdoub}) also suggests that the Mellin variables $s$ and $t$ are conjugate to the twist of the operators appearing in the $s$ and $t$-channel expansion respectively. 

The Mellin transform of the four point function thus has a simple pole structure -- the Mellin amplitude $A(s,t)$ contains simple poles corresponding to the physical operators of the theory, while the gamma functions in the measure contain poles that correspond to the double-trace primaries of the theory, 
\begin{align}
    \Delta = \Delta_i + \Delta_j + J + 2n.
\end{align}
As discussed near (\ref{dop}), apart from theories that are perturbatively close to generalised free fields, these operators do not exist in a \textit{generic} interacting theory. These operators need to eventually cancel in the full crossing symmetric calculation of the four-point function. In fact, this constraint is at the heart of the Mellin-Polyakov bootstrap program \cite{Gopakumar:2016cpb,Gopakumar:2018xqi,Polyakov:1974gs}. 

We will be interested in large $N$ theories, where these operators are actual composite primaries present in the spectrum.  In these theories, Mellin space representation is particularly useful since the information about the single-trace primaries encoded in the Mellin amplitude is nicely separated from the double-trace primaries encoded in the measure. This feature has been used to great advantage by Aharony et al. \cite{Aharony:2016dwx} to calculate the anomalous dimensions in large $N$ theories. We discuss this technique in appendix \ref{coll} and use it to double check some of our results for the anomalous dimensions of the double-trace primaries.

\subsection{A Mellin inversion formula}\label{meat}
The goal of this section is to express the inversion formula in Mellin space. In $d=2$ using the symmetry $z \leftrightarrow \zb$, the inversion formula (\ref{cinv}) becomes,
\begin{align}
c^t(\Delta,J)= \frac{ \kappa_{\Delta+J}}{2(1+\delta_{\Delta,1})} \int_0^1 \frac{dz d\bar z}{z^2 \bar z^2} \left((1-z)(1-\zb)\right)^{a+b}  k_{J-\Delta+2}(z) k_{\Delta+J}(\zb)  \text{dDisc}[\mathcal G]\, . \label{sinv}
\end{align}
Using the Mellin transform (\ref{Mellin}) of the four point function we obtain,
{\footnotesize
\begin{align}
c^t(\Delta,J)= & \frac{ \kappa_{\Delta+J}}{2 ( 1 + \delta_{\Delta,1})} \int_0^1 \frac{dz d\bar z}{z^2 \bar z^2}\int \frac{ds dt}{(2\pi i )^2}   k_{J-\Delta+2}(z) k_{\Delta+J}(\zb)  \text{dDisc}\left[(z \zb )^{\frac s 2}\left((1-z) (1-\zb) \right)^{\frac{t-\Delta_2 - \Delta_3}{2}}\right] \nonumber \\
& \left((1-z)(1-\zb)\right)^{a+b}   \Gamma \left(\frac{\Delta_1 + \Delta_4 -t}{2} \right) \Gamma \left(\frac{\Delta_2 + \Delta_3 -t}{2} \right) \Upsilon(s,t) A(s,t) \nonumber \\
= &  \frac{\kappa_{\Delta+J}}{( 1 + \delta_{\Delta,1})} \int \frac{ds dt}{(2\pi i )^2} \int_0^1 \frac{dz}{z^{2-\frac s 2}} k_{J-\Delta+2}(z) (1-z)^{\frac{t-\Delta_1 - \Delta_4}{2}}  \int_0^1 \frac{d\zb }{\zb^{2- \frac s 2}} k_{\Delta+J}(\zb) (1-\zb)^{\frac{t-\Delta_1 - \Delta_4}{2}}   \nonumber \\
& \times \frac{ \pi^2 \Upsilon(s,t)   A(s,t) }{\Gamma\left( \frac{2 + t - \Delta_2  - \Delta_3}{2}  \right) \Gamma\left( \frac{2 + t - \Delta_1  - \Delta_4}{2} \right)}. \label{minv2}
\end{align}}

\noindent The crucial step between the two lines above is the disappearance of the $t$-channel poles due to $\Gamma \left(\frac{\Delta_1 + \Delta_4 -t}{2} \right) \Gamma \left(\frac{\Delta_2 + \Delta_3 -t}{2} \right)$, and the $\dDisc_t$. The double discontinuity produces zeros 

{\footnotesize
\begin{align}
\frac{\text{dDisc}[(z \zb )^{\frac s 2}\left((1-z) (1-\zb) \right)^{\frac{t-\Delta_2 - \Delta_3}{2}}]}{(z \zb )^{\frac s 2}\left((1-z) (1-\zb) \right)^{\frac{t-\Delta_2 - \Delta_3}{2}}} = 2\sin\left( \frac \pi 2 \left ( t-\Delta_2 - \Delta_3 \right)\right) \sin\left( \frac \pi 2 \left (t-\Delta_1 - \Delta_4\right)\right), \label{dsin}
\end{align}}

\noindent that exactly cancel the poles of the gamma functions $\Gamma \left(\frac{\Delta_1 + \Delta_4 -t}{2} \right) \Gamma \left(\frac{\Delta_2 + \Delta_3 -t}{2} \right)$. In the last line of (\ref{minv2}) we have rewritten the product of the sine and the gamma function using the identity,
\begin{align}
\Gamma(-x) \sin (\pi x) = -\frac{\pi}{\Gamma(1+x)}. \label{sine}
\end{align}
The zeros of the gamma functions that got cancelled, would have precisely given rise to the double-trace operators of the theory
\begin{align}
    t= \Delta_1 + \Delta_4 + 2m , \qquad t= \Delta_2 + \Delta_3 + 2m,  \qquad m=0,1,2\ldots \, .
\end{align}
The fact that these zeroes cancel in our formula, is just a neat representation of the well known fact mentioned near (\ref{poleg}) -- the inversion formula gets no contribution from the double-trace primaries. To proceed we need to evaluate the integral of the type,
\begin{align}
I \equiv  \int_0^1 dx k_{h}(x)x^\alpha (1-x)^\beta.  \label{samQ}
\end{align}
Using the integral representation of $_{3}F_2$ we can evaluate the above integral to be,
{\small
\begin{align}
I  = \Gamma (\beta +1) \Gamma (2 h)  \Gamma (h+\alpha +1) & \, _3\tilde{F}_2(a+h,b+h,h+\alpha +1;2 h,h+\alpha +\beta +2;1),\nonumber \\
_{3} \tilde{F}_2 (a_1,a_2, a_3;b_1,b_2;z) & = \frac{_{3} F_2 (a_1,a_2, a_3;b_1,b_2;z)}{\Gamma(b_1) \Gamma(b_2)}. \label{samA}
\end{align}}

\noindent A nice feature of the regularized hypergeometric functions ${}_3\tilde F_2$ is that they have no poles for any finite values of their parameters $a_i$ and $b_i$. The above representation of the ${}_3F_2$ is valid only when, 
\begin{align}
\text{Re}(a+b-\beta )<1, \qquad \text{Re}\left(\alpha + \frac h 2 \right)>-1, \qquad \text{Re}(\beta )>-1 . \label{conv}
\end{align}
These inequalities arise from the boundaries of the integral (\ref{samQ}) near $x=0$ or $x=1$, and seem quite restraining. Indeed, we find that some of the cases of our interest do violate them. We will show how to avoid this issue using the analyticity of our result in the following section. Using all of this in (\ref{sinv}) we have for a CFT$_2$,

{\footnotesize
\begin{align}
c^t(\Delta,J)  = &  \frac{\pi^2 \Gamma (2-\tau ) \Gamma\left(\Delta+ J \right) \kappa_{\Delta+J} }{\left( 1+ \delta_{\Delta,1}\right) }  \int \frac{ds dt}{(2\pi i )^2}    \Gamma \left(\frac{s-\tau }{2}\right) \Gamma \left(\frac{s+\Delta +J -2}{2}\right) \frac{   \Gamma \left(\frac{2+t-\Delta_1-\Delta_4}{2}\right)}{\Gamma \left(\frac{2+t-\Delta_2-\Delta_3}{2}\right)} \nonumber \\
  & \pFq{3}{2}{\frac{2a+2-\tau }{2},\frac{s-\tau}{2},\frac{2b+2-\tau}{2}}{\frac{s+t-\Delta_1 -\Delta_4-\tau+2}{2},2-\tau}{1} \, \pFq{3}{2}{\frac{2a+\Delta +J }{2},\frac{s+\Delta+J -2}{2},\frac{2b+\Delta +J }{2}}{\frac{s+t+\Delta +J -\Delta_1-\Delta_4}{2},\Delta + J}{1} \, \Upsilon(s,t)  A(s,t), \label{Mellin2d}
\end{align}}

\noindent where $\tau=\Delta-J$. Due to (\ref{conv}) our result, for the moment, is valid only when
{\footnotesize
\begin{align}
\text{Re}(s)-\tau > 0, \quad \text{Re}(s)+\Delta+J>2, \quad \Delta_1 + \Delta_4 - 2 < \text{Re}(t), \quad \Delta_2 + \Delta_3 - 2 < \text{Re}(t). \label{2dval}
\end{align}}
\noindent Throughout this note we will only mention the $t$-channel contribution, since the $u$-channel contribution can be obtained by a simple substitution $\Delta_1 \leftrightarrow \Delta_2$. The analysis for four dimensions is identical and we only state the final result,

{\scriptsize
\begin{align}
c^t(\Delta,J)  = &  \pi^2 \Gamma (4-\tau ) \Gamma(\Delta+J) \kappa_{\Delta+J}  \int \frac{ds dt}{(2\pi i)^2}  \Gamma \left(\frac{J+s-\Delta }{2}\right) \Gamma \left(\frac{J+s+\Delta -4}{2}\right)   \frac{ \Gamma \left(\frac{t-\Delta_1-\Delta_4+4}{2}\right)}{ \Gamma \left(\frac{t-\Delta_2-\Delta_3+2}{2}\right)}  \nonumber \\
 &   \Bigg \{ \, \pFq{3}{2}{\frac{J-\Delta +2a + 4}{2},\frac{J+s-\Delta}{2},\frac{J-\Delta +2b +4}{2}}{J-\Delta +4,\frac{J+s+t-\Delta -\Delta_1-\Delta_4+4}{2}}{1} \,  \pFq{3}{2}{\frac{ J+\Delta+2a}{2},\frac{J+s+\Delta -4}{2},\frac{J+\Delta +2b}{2}}{J+\Delta,\frac{ J+s+t+\Delta -\Delta_1-\Delta_4-2}{2}}{1}  \nonumber \\
 &  -  \, \pFq{3}{2}{\frac{J-\Delta +2a +4}{2},\frac{J+s-\Delta}{2},\frac{J-\Delta +2b+4}{2}}{J-\Delta +4,\frac{J+s+t-\Delta -\Delta_1-\Delta_4+2}{2}}{1} \,  \pFq{3}{2}{ \frac{J+\Delta +2a}{2},\frac{J+s+\Delta -4}{2},\frac{J+\Delta +2b}{2}}{J+\Delta ,\frac{J+s+t+\Delta -\Delta_1-\Delta_4}{2}}{1} \Bigg \} \Upsilon(s,t) A(s,t) . \label{Mellin4d}
\end{align}}

\noindent Like before, our result for now, is only valid when
{\footnotesize
\begin{align}
\text{Re}(s)-\tau > 0, \quad \text{Re}(s)+\Delta+J>4, \quad \Delta_1 + \Delta_4 - 2 < \text{Re}(t), \quad \Delta_2 + \Delta_3 - 2 < \text{Re}(t). \label{4dval}
\end{align}}
We refer to (\ref{Mellin2d}) and (\ref{Mellin4d}) as the Mellin inversion formula, and is the main result of our note. We use it in the following sections to obtain the CFT data in large $N$ theories.

\subsection{Convergence in Mellin space} \label{melcon}
There are a couple of interesting points that we would like to discuss before we put our formula to use. It is well known that the inversion formula only works when $J>1$. This can be understood by looking at the Regge limit of (\ref{cinv}) -- $z,\zb \rightarrow 0$ (after taking $\zb$ around 1 via the $\dDisc_t$), where the integrand could possibly diverge. The growth of the integrand in turn relies on the behavior of the dDisc of the four-point function in the Regge limit. This limit has now been well understood \cite{Maldacena:2015waa,Hartman:2015lfa}, and it can be shown that $\dDisc[\mathcal G(z,\zb)]$ in any unitary CFT stays $O(1)$ in the Regge limit. Using this fact and 
\begin{align}
\lim_{\substack{z,\zb \rightarrow 0 \\ z \sim \zb}}g_{J+d-1,\Delta+1-d}(z,\zb) = z^{J+d-1} + \ldots, \quad
\lim_{\substack{z,\zb \rightarrow 0 \\ z \sim \zb}} \mu(z,\zb) = z^{-(d+2)}, \label{regexp}
\end{align}
in (\ref{cinv}), it can be easily checked that $J>1$ for the integral to converge. 

We are interested in inverting individual Witten diagrams. The Witten diagram corresponding to the exchange of an operator $O_k$ with dimensions $(\Delta_k,J_k)$ can be written as a linear combination of conformal blocks of dimension $(\Delta_k,J_k)$ and its shadow $(d-\Delta_k,J_k)$ \cite{Costa:2012cb,Gopakumar:2016cpb}. Since the individual conformal blocks in the Regge limit grow as $z^{1-J_k}$ \cite{Cornalba:2006xm,Li:2017lmh}, the integral (\ref{cinv})
\begin{align}
c^t(\Delta,J) \sim \int \! d^2z \, z^{-(d+2)} z^{J+d-1}z^{1-J_k},
\end{align}
converges only when, 
\begin{align}
J>J_k. \label{lowspin}
\end{align}
This will be important when we discuss the contact diagram in section \ref{seccon}. Note that in a large $N$ sparse CFT, where the graviton is the highest spin particle, (\ref{lowspin}) would mean that the inversion formula works only when $J>2$ (and not $J>1$). This is an artifact of the large $N$ expansion, and indeed at finite $N$, $J>1$ is sufficient \cite{Caron-Huot:2017vep,Simmons-Duffin:2017nub}. 

The reader might be also concerned that the Mellin inversion formula apparently has more constraints than the usual Lorentzian inversion formula, namely the inequalities in (\ref{2dval}) and (\ref{4dval}). These inequalities come from the $\zb \rightarrow 1 \lor z \rightarrow 1 \lor z \rightarrow 0 \lor \zb \rightarrow 0 $ limit in the integral (\ref{minv2}). These constraints are artificial, as was originally argued by Caron-Huot. One can define the integral by cutting off a small circle around the singularity and dropping the singular terms as the radius goes to zero. These terms can then be integrated back if the integral is analytic. Finding the right analytic continuation can be tricky sometimes, but fortunately the hypergeometric functions have been studied extensively in the mathematics literature. Using the following identity from \cite{Buhring:1987:BUA:36949.36951},
\begin{align}
\pFq{3}{2}{a,b,c}{e,f}{1} =\pFq{3}{2}{e-c,f-c,r}{r+a,r+b}{1} \, \frac{\Gamma (r)}{\Gamma (c)} \, ,  \label{3f2}
\end{align}
where $r=e+f-a-b-c$, we can analytically continue our results (\ref{Mellin2d}) and (\ref{Mellin4d}) to any value of $s$ and $t$. Recall that the ${}_3\tilde{F}_2(a_1,a_2,a_3;b_1,b_2;1)$ at unit argument is well defined only when $$\sum_{i=0}^2 b_i - \sum_{j=0}^3 a_j >0.$$ The beauty of (\ref{3f2}) lies in the fact that the right hand side always satisfies the above inequality when $c$ is positive. While for negative values of $c$, it can be checked that the gamma functions multiplying the ${}_3\tilde{F}_2$ take care of the singularities, if any. We use this identity in all of our following results, and find that we get correct finite answers even if the inequalities (\ref{2dval}) and (\ref{4dval}) are violated.

\section{Witten diagrams and anomalous dimensions \label{sec3}}
We will now put our formula to use in the context of AdS/CFT. 
We work with a bulk theory of scalar fields that have cubic or quartic interactions. In a large $N$ theory dual to AdS, the four-point function can be expressed as a sum over Witten diagrams, i.e. as a perturbative expansion in $1/N$.
\begin{figure}
    \centering
    \includegraphics[scale=0.82]{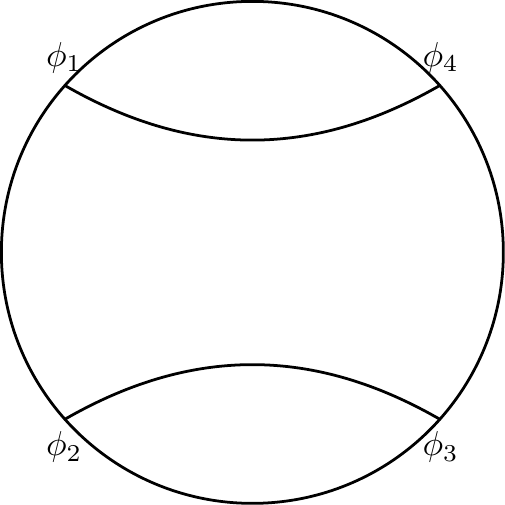}
    \caption{Generalised free fields in the $t$-channel $(23) \rightarrow (14)$.}
    \label{gff}
\end{figure}
Likewise, the inversion formula for $c^t(\Delta,J)$ can be organized in a large $N$ expansion, 
\begin{align}
    c^t_{\phi^3}(\Delta,J) &=  c^t_{gff}(\Delta,J) + c^t_{exchange}(\Delta,J)+c^t_{box}(\Delta,J) + \ldots \, , \\
    c^t_{\phi^4}(\Delta,J) &=  c^t_{gff}(\Delta,J) + c^t_{contact}(\Delta,J)+c^t_{bubble}(\Delta,J) + \ldots \, ,
\end{align}
where $gff$ stands for generalised free fields, and the other terms come from the respective diagrams. In both expansions the first term is $O(1)$, the second is $O(1/N^2)$ and the third is $O(1/N^4)$. We will use the Mellin amplitude of Witten diagrams in the Mellin inversion formula to evaluate the OPE coefficients and anomalous dimensions of leading-twist double-trace primaries in $d=2$ and $d=4$ boundary spacetime dimensions. The Mellin amplitudes of Witten diagrams are remarkably simple and have been calculated in the literature since long, starting with the seminal work by Penedones \cite{Penedones:2010ue,Fitzpatrick:2011ia,Yuan:2018qva,Cardona:2017tsw}.

In section \ref{seccon}, we evaluate the contribution of the contact Witten diagram to $c^t(\Delta,J)$. In section \ref{secexc} and \ref{secbub}, we evaluate the contribution of the exchange and bubble Witten diagrams to $c^t(\Delta,J)$. As mentioned in the introduction, $c^t(\Delta,J)$ has poles corresponding to the double-trace primaries of the theory, and the exchange and bubble Witten diagrams give the corresponding CFT data at $O(1/N^2)$ and $O(1/N^4)$ respectively. Like before, we only discuss $c^t(\Delta,J)$ since $c^u(\Delta,J)$ can be obtained simply by the substitution $\Delta_1 \leftrightarrow \Delta_2$. 

\subsection{Contact diagram} \label{seccon}
\begin{figure}
\centering
\includegraphics[scale=0.72]{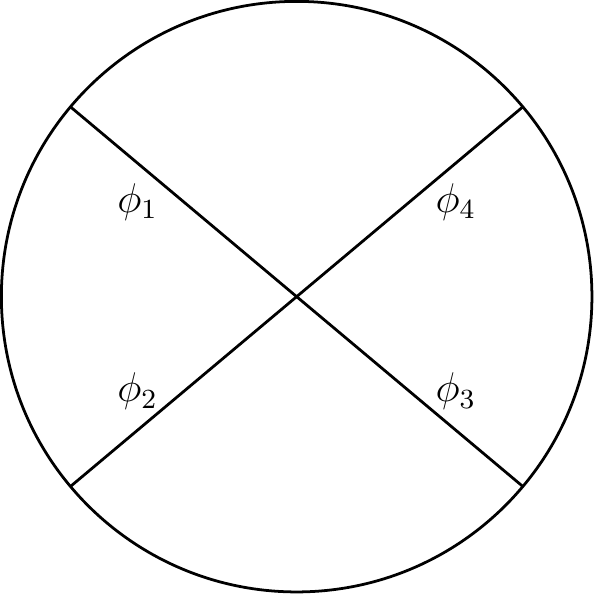}    
\caption{Four-point contact Witten diagram.}
\label{contactd}
\end{figure}
\begin{figure}
    \centering
    \includegraphics[scale=1]{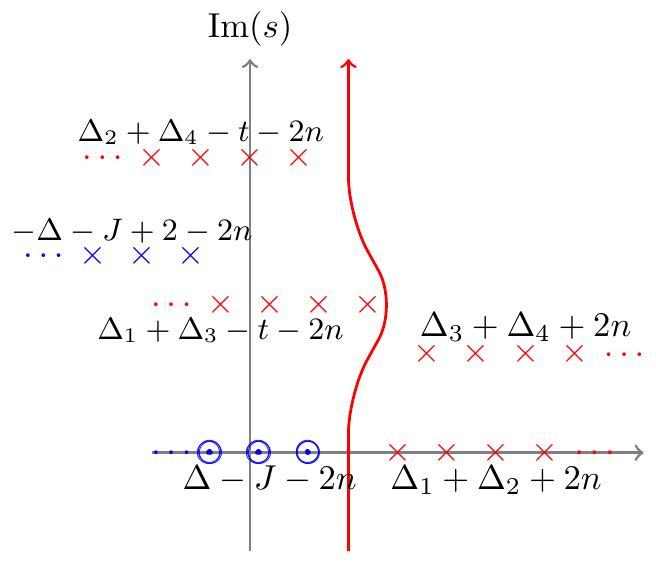} \qquad
    \includegraphics[scale=1]{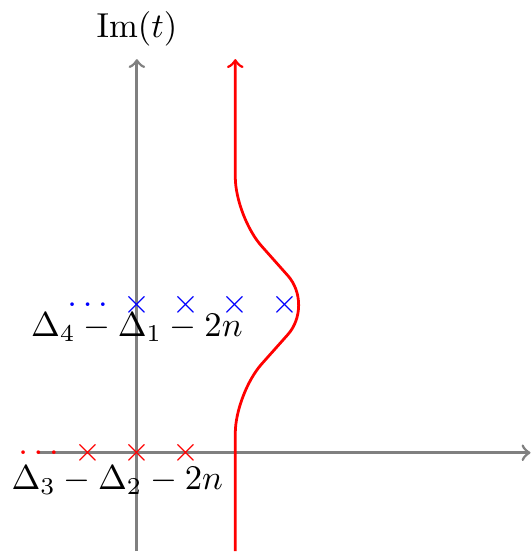}
    \caption{Poles in the complex $s$ and $t$ planes. We have complexified the scaling dimensions here for the purposes of clarity.}
    \label{supoles}
\end{figure}

We start with the four-point contact diagram of figure \ref{contactd} with the interaction vertex $ \lambda \phi_1 \phi_2 \phi_3 \phi_4$. The Mellin amplitude of this diagram is simply a constant \cite{Penedones:2010ue}:
\begin{align}
A(s,t) = \mathcal N \lambda R^{3-d} . \label{melcont}
\end{align}
Here $R$ is the AdS radius and
\begin{align}
\mathcal N \equiv \frac{\pi^{d/2}}{2}\Gamma \left(  \frac{\sum_{i=1}^n \Delta_i - \frac d 2}{2} \right) \prod_{i=1}^n \frac{1}{2\pi^{d/2} \Gamma(\Delta_i- \frac d 2+1)}. \label{norm}
\end{align}
Let us study the analytic structure of the integrand in the Mellin inversion formula (\ref{Mellin2d}) for the Mellin amplitude (\ref{melcont}). Since the ${}_3\tilde{F}_2$ defined in (\ref{samA}) is analytic for any finite value of its parameters, the only singularities in $s$ and $t$ come from the gamma functions,
\begin{align}
     & \frac{\Gamma\left( \frac{s - \Delta + J}{2}  \right)\Gamma\left( \frac{s + \Delta + J  - 2}{2}  \right) \Gamma\left( \frac{2+t- \Delta_1 - \Delta_4}{2}\right)\Upsilon(s,t)}{\Gamma\left( \frac{2+t-\Delta_2 - \Delta_3}{2} \right)} . \nonumber 
\end{align}
We reproduce the definition of $\Upsilon(s,t)$ for the convenience of the reader, 
{\small
\begin{align}
    \Upsilon(s,t) = \Gamma\left(\frac{\Delta_1 + \Delta_2 - s}{2}\right)\Gamma\left(\frac{\Delta_3 + \Delta_4 - s}{2}\right)\Gamma\left(\frac{s+t-\Delta_2 - \Delta_4}{2}\right)\Gamma\left(\frac{s+t-\Delta_1  -\Delta_3}{2}\right). \nonumber
\end{align}}
The double-trace poles of $\Upsilon(s,t)$ are shown in red crosses in the left panel of figure \ref{supoles}, while the poles that arise from the $(z,\zb)$ integral of the inversion formula are shown in blue. The contour is closed to the right where it picks up the double-trace poles at $s=\Delta_1 + \Delta_2 + 2n $ and $s= \Delta_3 + \Delta_4 + 2n$. In fact, the pinching of the integration contour between these double-trace poles in $s$ and the poles $s=\Delta-J+2n$ (shown in blue circles in the left panel of figure \ref{supoles}) gives the eventual pole in the inversion formula at $\Delta=\Delta_1 + \Delta_2 + J + 2n$ and $\Delta=\Delta_3 + \Delta_4 + J + 2n$.

The remaining integrand only has poles in $t$ that are entirely on the left hand side of the $t$ integration contour, as shown in the right panel of figure \ref{supoles}. Thus the contour can be closed trivially to the right (after making sure that the arcs at infinity can be safely dropped) to obtain,
\begin{align}
c^t_{contact}(\Delta,J) = 0 . \label{contact}
\end{align}
More qualitatively we observe that -- since the $\dDisc_t$ kills the $t$-channel double-trace gamma functions in the Mellin inversion formula, the only way the contour integrals can not vanish is if the Mellin amplitude has a pole or a branch cut in $t$. Since the Mellin amplitude for the contact diagram is just a constant, (\ref{contact}) is obvious. Similar arguments hold for the $u$ channel.

But this definitely cannot be the right answer, since the contact diagrams can be decomposed in the $s$ channel, and so the $s$ channel coefficients (the $a^{12}_{n,J}$ and  $a^{34}_{n,J}$ below) obviously cannot be zero. To understand what's going on we need to first understand what operators appear in the $s$-channel conformal block decomposition of contact Witten diagrams \cite{Zhou:2018sfz},
\begin{align}
    W_{contact}(x_i)= \sum_{J=0}^{J_{max}} \sum_{n=0}^{\infty} \left( a^{12}_{n,J} g_{\Delta_1 + \Delta_2 + 2n + J,J} (x_i) +a^{34}_{n,J} g_{\Delta_3 + \Delta_4 + 2n + J,J} (x_i) \right).
\end{align}
Here $J_{max}$ is the spin of the contact interaction i.e. zero. We see that the $s$-channel decomposition only contains operators with spin $J=0$. But following the discussion near (\ref{lowspin}) about the convergence of the inversion formula, we know that the Mellin inversion formula for the contact diagram is valid only when $J > 0$. This is an intrinsic limitation of the inversion formula that stops us from getting the CFT data for low enough spins. Thus ultimately (\ref{contact}) is the wrong answer, but we had expected it to be so. Similar arguments hold for contact interactions with derivatives.

\subsection{Exchange diagram}  \label{secexc}
\begin{figure}
    \centering
    \includegraphics[scale=0.85]{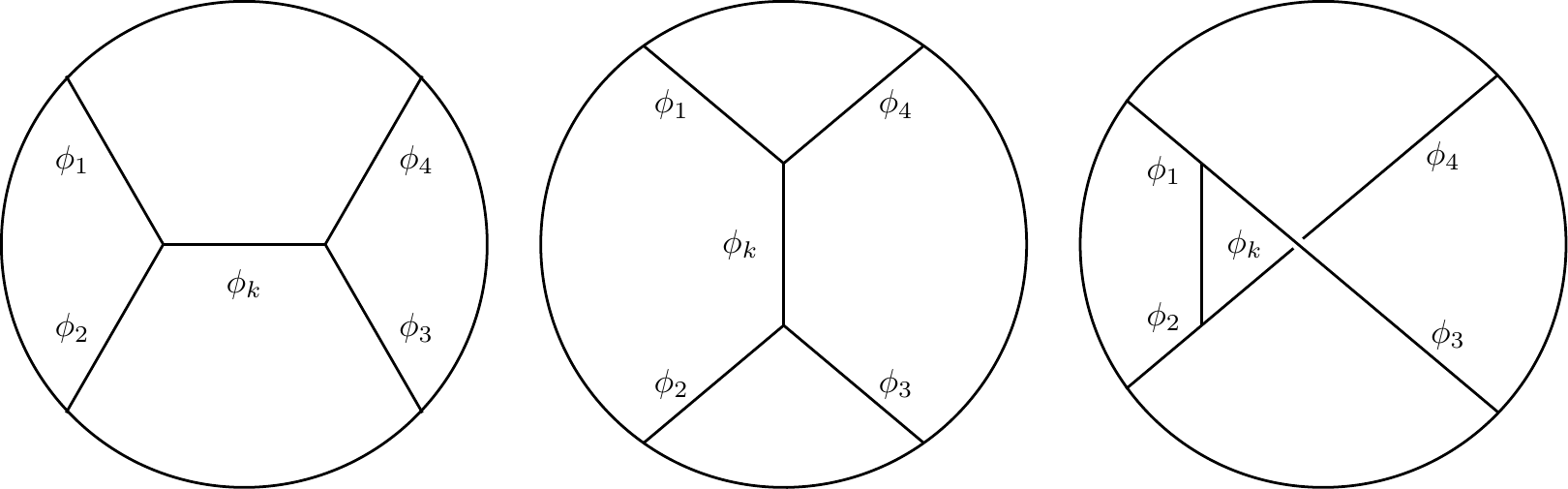}
    \caption{$s,t$ and $u$ scalar exchange Witten diagrams.}
    \label{stu}
\end{figure}

We now proceed to evaluate the CFT data for the bulk scalar theory with cubic vertices. The first nontrivial contribution to the four-point function after the generalised free field theory contribution of figure \ref{gff}, are the exchange diagrams of figure \ref{stu}. The exchange of an operator $\phi_{k}$ at tree level is given by the sum of three Witten diagrams,
\begin{align}
W_{\phi_k}=W^s_{\phi_k} +W^t_{\phi_k} +W^u_{\phi_k} ,
\end{align}
where $s,t$ and $u$ correspond to $(12)\rightarrow(34), (14)\rightarrow(23) $ and $(13)\rightarrow(24)$ respectively and $\Delta_k$ is the scaling dimension of the exchanged operator $\phi_k$. Let us start with the 
Mellin amplitude of the $t$-channel scalar exchange Witten diagram \cite{Penedones:2010ue},
\begin{align}
A(s,t)  = -\mathcal N &  g^2 R^{5-d} \sum_{m=0}^{\infty} \frac{R_m}{t - \Delta_k - 2m}, \label{melexc} 
\end{align}
 where $\mathcal N$ was defined in (\ref{norm}) and $g$ is the bulk coupling constant. The residue is given by,
\begin{align}
R_m  =  \frac{\Gamma\left(  \frac{\Delta_1 + \Delta_4 + \Delta_k - d}{2} \right)\Gamma\left(  \frac{\Delta_2 + \Delta_3 + \Delta_k - d}{2} \right)  }{2\Gamma(\frac{\Delta_1 + \Delta_2 +\Delta_3 + \Delta_4 - d}{2})}  & \frac{\left( 1 + \frac{\Delta_k-\Delta_1 - \Delta_4}{2}  \right)_m\left( 1 + \frac{\Delta_k-\Delta_3 - \Delta_2}{2}  \right)_m }{m! \Gamma\left(\Delta_k - \frac{d}{2}+1 +m \right)}, \label{res}
\end{align}
\noindent and has already been evaluated at the pole $t= \Delta_k + 2m$. 

The $s$ and $u$-Witten diagrams do not contribute to $c^t(\Delta,J)$. As argued near (\ref{melcont}), the inversion formula will vanish unless the Mellin amplitude has a pole or a cut in $t$. Since the $s$ and $u$-Witten diagrams only have poles in $s$ and $u$ respectively i.e. are analytic in $t$, their contribution vanishes. Similarly only the $u$-Witten diagram contributes to $c^u(\Delta,J)$. The Mellin amplitude (\ref{melexc}) has a pole at $t = \Delta_k$ in addition to an infinite number of poles labelled by $m$. These are called satellite poles, and schematically correspond to the descendants of $\phi_k$.

Using the Mellin amplitude (\ref{melexc}) in the inversion formula (\ref{Mellin2d}), we find that the pole structure of the complex $s$ plane is identical to that of the contact diagram. Like before the contour is closed to the right where it picks up the poles at 
\begin{align}
s = \Delta_1 + \Delta_2 + 2n \quad \text{or} \quad s = \Delta_3 + \Delta_4 + 2n,  \qquad n=0,1,2,\ldots \, . \label{spole}
\end{align}
\begin{figure}
    \centering
    \includegraphics[scale=1]{images/spole.pdf} \qquad
    \includegraphics[scale=1]{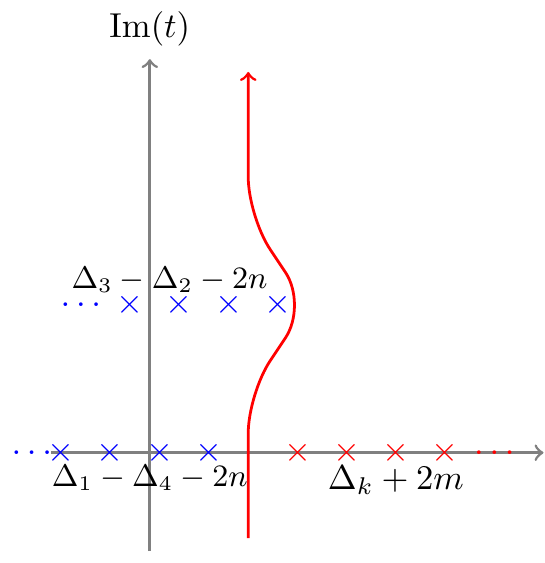}
    \caption{Complex $s$ and $t$ plane.}
    \label{stplane}
\end{figure}
However, the complex $t$ plane is different from that of the contact diagram. As shown in figure \ref{stplane}, there are poles on both sides of the contour, and it cannot be closed trivially. We close the contour to the right and pick the poles at,
\begin{align}
    t= \Delta_k + 2m. \label{tpole}
\end{align}
Using all of this in the inversion formula (\ref{Mellin2d}) we find that it has poles at
\begin{align}
    \Delta= \Delta_1 + \Delta_2 + J + 2n, \qquad \Delta=\Delta_3 + \Delta_4 + J + 2n, \qquad  n=0,1,2\ldots \, .
\end{align}
As mentioned before, these are precisely the operator dimensions of the double-trace primaries in the $s$-channel. The residue of the inversion formula at these poles then gives the OPE coefficient of these operators. Focusing on the $\Delta_1 + \Delta_2 + J + 2n $ primary we have,
\begin{align}
c^t_{exchange}(\Delta,J) \supset \frac{a_{n,J}}{\Delta - (\Delta_1 + \Delta_2 + J + 2n + \gamma)}, \qquad n=0,1,2\ldots \, . \label{struc}
\end{align}
Writing the CFT data perturbatively in $\frac 1 N$,
\begin{align}
a_{n,J} = a^{(0)}_{n,J}  + \frac{1}{N^2}a^{(1)}_{n,J}  + \frac{1}{N^4}a^{(2)}_{n,J} + \ldots \, , \qquad \gamma_{n,J} =  \frac{1}{N^2}\gamma^{(1)}_{n,J}  + \frac{1}{N^4}\gamma^{(2)}_{n,J} + \ldots \, , \label{exp}
\end{align}
the OPE function (\ref{struc}) becomes,
{\scriptsize
\begin{align}
    c^t(\Delta,J) \supset & \frac{a^{(0)}_{n,J}}{\Delta -\Delta_1 - \Delta_2 -J-2n} + \frac{1}{N^2} \left(\frac{a^{(0)}_{n,J} \gamma^{(1)}_{n,J}}{(\Delta -\Delta_1 - \Delta_2-J-2n)^2}+\frac{a^{(1)}_{n,J}}{\Delta -\Delta_1 - \Delta_2-J-2n}\right) \nonumber \\
    & + \frac{1}{N^4} \left(\frac{a^{(0)}_{n,J} \gamma^{(2)}_{n,J}+a^{(1)}_{n,J}\gamma^{(1)}_{n,J} }{(\Delta -\Delta_1 - \Delta_2-J-2n)^2}+\frac{a^{(0)}_{n,J} \left(\gamma^{(1)}_{n,J}\right)^2}{(\Delta -\Delta_1 - \Delta_2-J-2n)^3}+\frac{a^{(2)}_{n,J}}{\Delta -\Delta_1 - \Delta_2-J-2n}\right) + \ldots \, . \label{cstruc}
\end{align} }

\noindent The zeroth order coefficients $a^{(0)}_{n,J}$ can be obtained using generalised free field theory \cite{Heemskerk:2009pn}. For non-identical external operators they vanish. This is because $gff$ correlators are just given by Wick contractions, and the two point function $\braket{\phi_i \phi_j}$ vanishes unless $i=j$. The above then becomes,

{\scriptsize
\begin{align}
     c^t(\Delta,J) = &  \frac{1}{N^2} \frac{a^{(1)}_{n,J}}{\Delta - \Delta_1 - \Delta_2-J-2n} + \frac{1}{N^4} \left(\frac{a^{(1)}_{n,J}\gamma^{(1)}_{n,J} }{(\Delta - \Delta_1-\Delta_2-J-2n)^2}+\frac{a^{(2)}_{n,J}}{\Delta - \Delta_1-\Delta_2-J-2n}\right) + \ldots \, .
\end{align}}

\noindent The tree level exchange diagram precisely provides the $a^{(1)}_{n,J}$. Using (\ref{Mellin2d}), and (\ref{melexc}) through (\ref{exp}) we have,

{\footnotesize
\begin{align}
    a^{(1)}_{0,J} = &  \frac{ 2 \pi^2 \kappa_{\Delta_1 + \Delta_2 + 2J} \Gamma (\Delta_k)^2 \Gamma (J+\Delta_1+\Delta_2-1) \Gamma (2 J+\Delta_1+\Delta_2) \Gamma \left(\frac{ \Delta_3-\Delta_2+\Delta_4-\Delta_1}{2}\right)}{ \Gamma \left(\frac{ \Delta_4-\Delta_1+\Delta_k}{2}\right) \Gamma \left(\frac{ \Delta_3 - \Delta_2+\Delta_k}{2}\right) \Gamma \left(\frac{2 J+\Delta_1+\Delta_2+\Delta_3-\Delta_4}{2}\right)  } \nonumber \\
    & \sum_{m=0}^\infty\frac{\left(\frac{ \Delta_1-\Delta_4+\Delta_k}{2}\right)_m \left(\frac{\Delta_2-\Delta_3+\Delta_k}{2}\right)_m}{m! \Gamma (m+\Delta_k)\Gamma\left( \frac{\Delta_1 + \Delta_4 - \Delta_k - 2m }{2}\right) \Gamma\left( \frac{\Delta_2 + \Delta_3 - \Delta_k - 2m }{2}\right) } \times \nonumber \\
    & \pFq{3}{2}{ \frac{2 J+\Delta_1+\Delta_2-\Delta_3+\Delta_4}{2} \quad ,\frac{2 m-\Delta_2-\Delta_3+\Delta_k+2}{2} \quad ,\frac{2 m+\Delta_2-\Delta_3+\Delta_k}{2}  }{\frac{2 J+2 m+\Delta_2-\Delta_3+\Delta_k+2}{2} \quad ,\frac{2 J+2 m+2 \Delta_1+\Delta_2-\Delta_3+\Delta_k}{2}}{1}. \label{ope2d}
\end{align}}

\noindent The calculation is straightforward, except one technical detail that is worth mentioning. If we recall (\ref{2dval}), it seems that (\ref{spole}) and (\ref{tpole}) will typically violate the regime of validity of the integral representation of the ${}_3 \tilde{F}_2$. As mentioned in section \ref{melcon} this is related to the fact that the inversion formula apparently blows up for several limiting values of $z,\zb$. This issue can be easily circumvented by using the result (\ref{3f2}). In writing (\ref{2dval}) (and all the results in the follow sections) we have already used this continuation, and the formula is valid for all positive values of $\Delta_i,J$ and $\Delta_k$. The analysis for the pole at $\Delta=\Delta_3 + \Delta_4 + J +2n$ is identical. The absence of the bulk coupling constant $g$ and AdS radius $R$ in (\ref{ope2d}) is an artifact of our normalisation that we elaborate in appendix \ref{normwitt}. 

The expression (\ref{ope2d}) looks rather unfortunate due to the sum over infinitely many satellite poles. We have not been able to resum them analytically. Although, the contribution of satellite poles falls exponentially in $m$ and it is possible to evaluate them numerically. For specific values of $\Delta_k$ the satellite poles are known to truncate, and we can obtain closed form results.  When 
\begin{align}
\Delta_k = \Delta_2 + \Delta_3  - 2m^\prime  \quad \text{or} \quad  \Delta_k = \Delta_1 + \Delta_4  - 2m^\prime,  \qquad \quad m^\prime=1,2,3,\ldots \, ,
\end{align}
the satellite poles truncate for all $m\geq m'$. This can be explicitly checked by evaluating the residue $R_m$, which vanishes for all $m \geq m'$. In the simplest case of $\Delta_k = \Delta_2 + \Delta_3  - 2 $ i.e. $m'=1$, there are no satellite poles and the OPE coefficient simply becomes,

{\footnotesize
\begin{align}
    a^{(1)}_{0,J} & = \frac{\Gamma (\Delta_2+\Delta_3-2) \Gamma (J+\Delta_1) \Gamma \left(\frac{\Delta_3+\Delta_4-\Delta_1-\Delta_2}{2}\right) \Gamma \left(\frac{2 J+\Delta_1+\Delta_2-\Delta_3+\Delta_4}{2}\right)}{\Gamma (\Delta_3-1) \Gamma (2 J+\Delta_1+\Delta_2-1) \Gamma \left(\frac{\Delta_1-\Delta_2-\Delta_3+\Delta_4+2}{2}\right) \Gamma \left(\frac{\Delta_2-\Delta_1+\Delta_3+\Delta_4-2}{2}\right)}.
\end{align}}

\noindent Closed form expressions for higher values of $m'$ can be similarly obtained. 

Another interesting exercise is to obtain the large $J$ behavior of our result (\ref{ope2d}). It can be shown that the contribution of a satellite pole $m=k$ is smaller by a factor of $J^{2k}$ than the leading term in (\ref{ope2d}) in the large $J$ limit. Using the following result of the hypergeometric function in (\ref{ope2d}),
\begin{align}
    \, _3F_2(a_1,a_2,a_3+J;b_1+J,b_2+J;1)= \sum_{n=0}^{k-1}\frac{(a_1)_n (a_2)_n (a_3+J)_n}{(b_1+J)_n (b_2+J)_n} \frac{1}{n!} + O\left( \frac{1}{J^k} \right), \label{3f2asymp}
\end{align}
the $m=0$ term gives the leading contribution,
{\scriptsize
\begin{align}
    a^{(1)}_{0,J \gg 1} =  \frac{4\sqrt{\pi } \Gamma (\Delta_k)  \Gamma \left(\frac{\Delta_3+\Delta_4-\Delta_1-\Delta_2}{2}\right) J^{\frac{ \sum_i \Delta_i-2 \Delta_k}{2}}}{2^{\Delta_1+\Delta_2+2 J}\Gamma \left(\frac{\Delta_4+\Delta_1-\Delta_k}{2}\right) \Gamma \left(\frac{\Delta_4-\Delta_1+\Delta_k}{2}\right) \Gamma \left(\frac{\Delta_2+\Delta_3-\Delta_k}{2}\right) \Gamma \left(\frac{\Delta_3-\Delta_2+\Delta_k}{2}\right)}\left(\frac{1}{J^{3/2}} +  O\left(\frac{1}{J^{5/2}}\right)\right). \label{largeJ}
\end{align}}
Subleading contributions can be obtained by including more satellite poles.

\subsubsection{Anomalous dimensions \texorpdfstring{$\frac{1}{N^2}\gamma^{(1)}_{0,J}$}{N}}
When the external scalars are identical $\Delta_i = \Delta_\phi$, the inversion formula has a double pole at, 
\begin{align}
    \Delta = 2\Delta_\phi + J + 2n. \label{double}
\end{align}
The mean field theory coefficients $a^{(0)}_{n,J}$ do not vanish for identical scalars, and are given by \cite{Heemskerk:2009pn}, 
\begin{align} 
    a^{(0)}_{n,J} = \frac{\Gamma (J+\Delta_\phi)^2 \Gamma (J+2 \Delta_\phi-1)}{J! \Gamma (\Delta_\phi)^2 \Gamma (2 J+2 \Delta_\phi-1)}. \label{mft}
\end{align}
The perturbative expansion of the OPE function (\ref{cstruc}) becomes,
{\footnotesize
\begin{align}
    c^t(\Delta,J) = &  \frac{a^{(0)}_{n,J}}{\Delta -2 \Delta_\phi-J-2n} + \frac{1}{N^2} \left(\frac{a^{(0)}_{n,J} \gamma^{(1)}_{n,J}}{(\Delta -2 \Delta_\phi-J-2n)^2}+\frac{a^{(1)}_{n,J}}{\Delta -2 \Delta_\phi-J-2n}\right) + \ldots \label{nexp}
\end{align}}
\noindent Thus the residue at the double-pole captures the anomalous dimensions of the double-trace primaries. Using the Mellin amplitude (\ref{melexc}) for identical scalars in the Mellin inversion formula (\ref{Mellin2d}), and evaluating the contour integrals at (\ref{spole}) and (\ref{tpole}) we find, 

{\scriptsize
\begin{align}
    \gamma^{(1)}_{0,J}  = & -  \sum_{m=0}^{\infty} \frac{2^{2\Delta_k} J! \Gamma \left(\frac{\Delta_k+1}{2}\right)^2 \Gamma (\Delta_\phi)^2  \Gamma (J+\Delta_\phi)    \left(\frac{\Delta_k}{2}\right)_m^2 }{ 2 \pi m!  \Gamma (m+\Delta_k) \Gamma \left(\Delta_\phi-\frac{\Delta_k}{2}-m\right)^2 } 
     \pFq{3}{2}{m+\frac{\Delta_k}{2},m+\frac{\Delta_k}{2}-\Delta_\phi+1,J+\Delta_\phi}{J+m+\frac{\Delta_k}{2}+1,J+m+\frac{\Delta_k}{2}+\Delta_\phi}{1}. \label{gamma}
\end{align}}

\noindent As before, the sum over $m$ corresponds to the sum over the satellite poles. For identical external scalars the $u$-channel contribution is identical, up to the explicit factor of $(-1)^J$ in the inversion formula (\ref{t+u}). Thus the final answer vanishes for odd spins, and is twice the above result for even. 

We can obtain closed form expressions when the satellite poles do truncate i.e. $2\Delta_\phi - \Delta_k = 2 m'$, where $m'$ is a positive integer. For the simplest case $2\Delta_\phi - \Delta_k = 2$ i.e. no satellite poles we simply find, 
\begin{align} 
\gamma^{(1)}_{0,J} & = - \frac{(\Delta_\phi-1) \Gamma (2 \Delta_\phi-1) \Gamma (J+1)}{2\Gamma (2 \Delta_\phi+J-1)}.
\end{align}
 Similar results can be obtained for other values of $m'$. We find that all of them match exactly with Liu et al. \cite{Liu:2018jhs}.  
\begin{figure}
    \centering
    \includegraphics[scale=0.292]{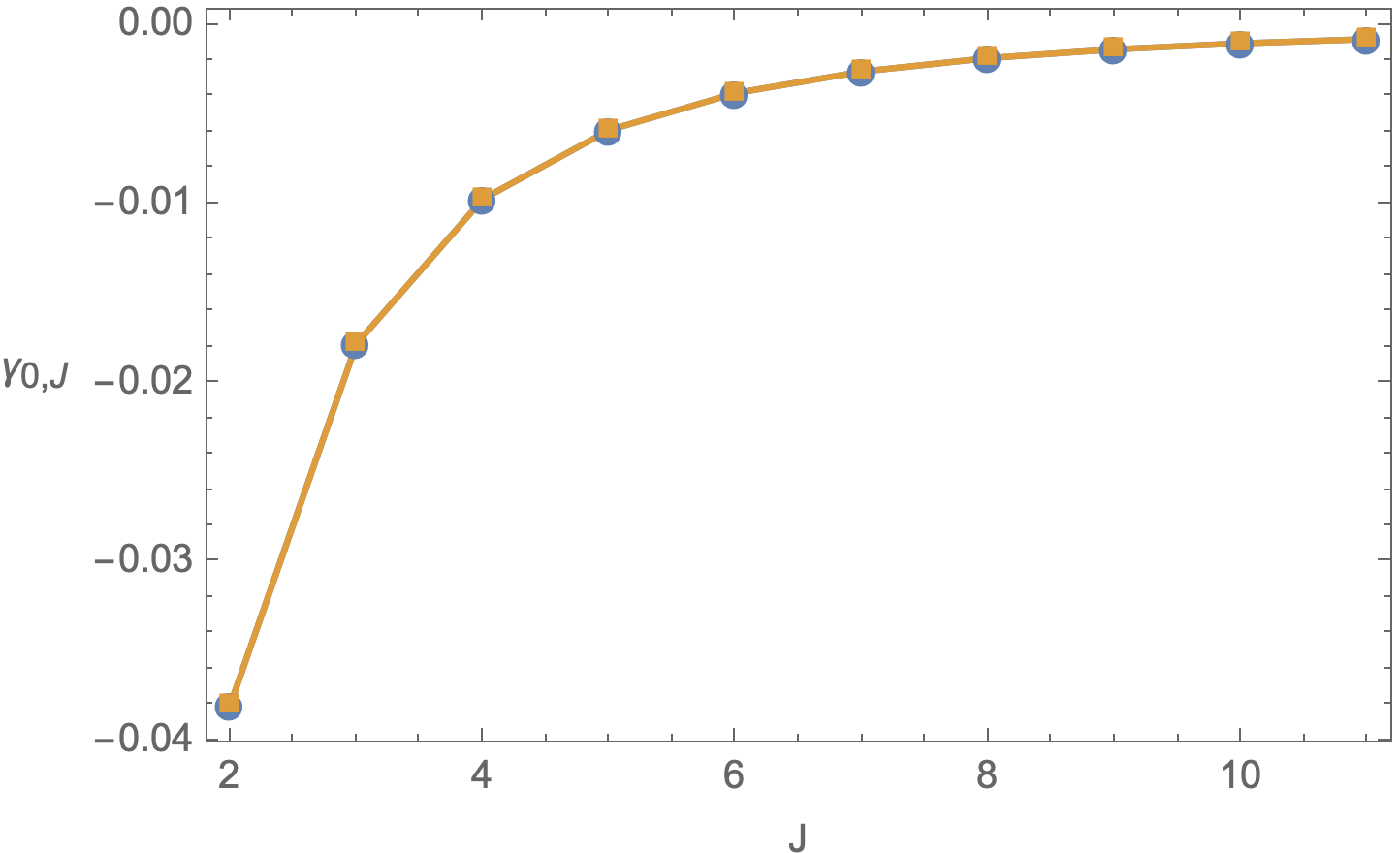} \,  \includegraphics[scale=0.3]{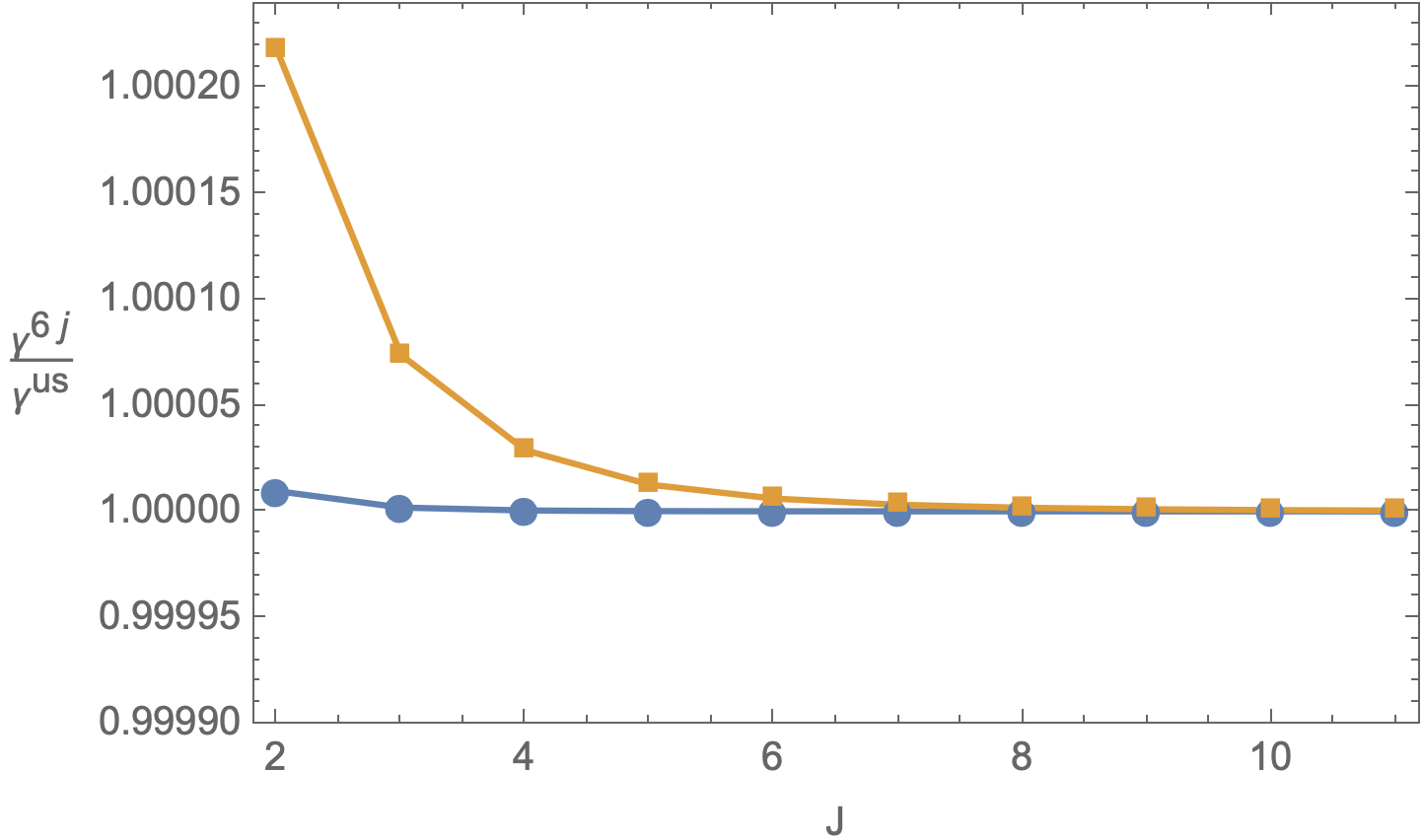}
    \caption{The left panel is a plot of $\gamma^{(1),us}_{0,J}$ and $\gamma_{0,J}^{6j}$ when the satellite poles do not truncate for e.g. $\Delta_\phi=2,\Delta_k=3$. We find that our result, even when we sum over just the first 2 satellite poles is indistinguishable from $\gamma^{6j}$. On the right panel we plot the ratio $\frac{\gamma^{6j}}{\gamma^{us}}$ for the case when we include the first 2 (yellow) and first 5 (blue) satellite poles. As expected, the blue points are closer to 1 than yellow. Same holds true for other values of $\Delta_\phi$ and $\Delta_k$.}
    \label{dsdme}
\end{figure}

When the satellite poles do not truncate i.e. $2\Delta_\phi - \Delta_k \not\in 2\mathbb Z^+$, there are three ways of checking our result. The first is to numerically compare our results with those of \cite{Liu:2018jhs} (which we refer as $\gamma^{6j}_{0,J}$). Since the contribution of the satellite poles fall exponentially in $m$, this is an easy check in Mathematica. As can be see in figure \ref{dsdme}, even summing up the first few satellite poles gives an almost indistinguishable match (less than $0.02\%$ error). The second is to check our result with the collinear decomposition of the four-point function in Mellin space\footnote{We thank Xinan Zhou for suggesting this.}. This method has been elaborated at several places, most notably in \cite{Aharony:2016dwx}. We discuss this technique in appendix \ref{coll} and show that our results match (satellite) pole by (satellite) pole with the results from collinear block decomposition. 

Finally, we could also look at the anomalous dimensions in the limit of asymptotically large $J$. Using (\ref{3f2}) we obtain,
{\scriptsize
\begin{align}
\gamma^{(1)}_{0,J} = -\frac{4^{\Delta_k} \Gamma \left(\frac{\Delta_k+1}{2}\right)^2 \Gamma (\Delta_\phi)^2}{4\pi  \Gamma (\Delta_k) \Gamma \left(\Delta_\phi-\frac{\Delta_k}{2}\right)^2} \left(\frac{2}{J^{\Delta_k}}-\frac{\Delta_k (2 \Delta_\phi-1)}{J^{\Delta_k+1}} -\frac{\Delta_k \left(\Delta_k^2-6 \Delta_k (2 \Delta_\phi-1)^2-4 (3 \Delta_\phi-2)^2\right)}{24 J^{\Delta_k+2}} + \ldots \right).
\end{align}}
The leading order term matches exactly with the result from large spin perturbation theory \cite{Fitzpatrick:2012yx,Komargodski:2012ek}. We can also express our results in terms of the conformal spin\footnote{We thank Eric Perlmutter for suggesting this to us}, $\tilde J \equiv (\Delta_\phi+J+n)(\Delta_\phi+J+n-1)$,
\begin{align}
    \gamma^{(1)}_{0,J} = -\frac{4^{\Delta_k} \Gamma \left(\frac{\Delta_k+1}{2}\right)^2 \Gamma (\Delta_\phi)^2}{4\pi  \Gamma (\Delta_k) \Gamma \left(\Delta_\phi-\frac{\Delta_k}{2}\right)^2} \left(\frac{2}{\tilde J^{\Delta_k}}-\frac{\Delta_k \left(\Delta_k^2-12 (\Delta_\phi-2) \Delta_\phi-4\right)}{24 \tilde J^{\Delta_k + 2}} + \ldots \right).
\end{align}
As argued in \cite{Alday:2015ewa,Alday:2015eya}, only even powers of the conformal spin enter in the large $\tilde{J}$ asymptotics.

\subsection{Bubble diagram} \label{secbub}
We now return to the bulk scalar theory with cubic interactions to calculate the OPE coefficients and anomalous dimensions of the double-trace primaries at $O\left(\frac{1}{N^4}\right)$. To our knowledge this has only been calculated recently by Aharony et al. \cite{Aharony:2016dwx} (cf. \cite{Ghosh:2018bgd}).
\begin{figure}
    \centering
    \includegraphics[scale=0.83]{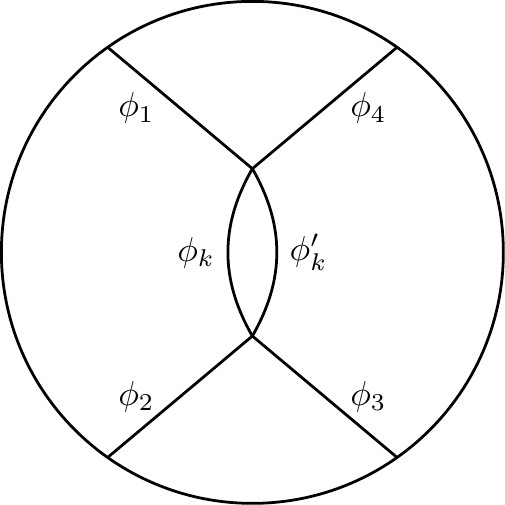}
    \caption{Bubble Witten diagram}
    \label{bubble}
\end{figure}
\noindent For non identical external scalars, $a^{(0)}_{n,J}$ vanishes and the order $\frac{1}{N^4}$ part of (\ref{cstruc}) becomes,
\begin{align}
    c(\Delta,J) \supset   \frac{1}{N^4} \left(\frac{a^{(2)}_{n,J}}{\Delta -\Delta_1 - \Delta_2-J-2n}\right). 
\end{align} 
\noindent The bubble Witten diagram of figure \ref{bubble}, provides us the OPE coefficient $a^{(2)}_{n,J}$ of the double-trace primaries. The Mellin amplitude of the $t$-channel bubble diagram can be found in \cite{Penedones:2010ue}. After evaluating the contour integrals therein we find,
\begin{align}
    M(t) = \mathcal N \lambda^2 R^{6-2d}\sum_{m_1,m_2} \frac{R_{m_1,m_2}}{t- (\Delta_k + \Delta'_k + 2m_1 + 2m_2) }. \nonumber
\end{align}
The amplitude has two sets of satellite poles, labelled by $m_1$ and $m_2$. As the form of the expression suggests they can be repackaged it into a single set,
\begin{align}
    M(t) & = \mathcal N \lambda^2 R^{6-2d}\sum_{m} \frac{R_m}{t- (\Delta_k + \Delta'_k + 2m) }, \qquad R_m   = \sum_{p=0}^{m}  R_{p,m-p}. \label{loopmel}
\end{align}
$\mathcal N$ was defined in (\ref{norm}) and the residue is given by,
{\tiny
\begin{align}
R_{p,m-p}  = & \frac{ - \Gamma \left(\frac{\Delta_1+\Delta_4-\Delta_k-\Delta'_k -2p}{2}\right) \Gamma \left(\frac{2 p+\Delta_1+\Delta_4+\Delta_k+\Delta'_k-2}{2}\right)  \Gamma \left(\frac{\Delta_2+\Delta_3-\Delta_k-\Delta'_k-2 p}{2}\right) \Gamma \left(\frac{2 p+\Delta_2+\Delta_3+\Delta_k+\Delta'_k-2}{2}\right)  }{4 \pi \Gamma (m-p+1)  \Gamma (\Delta_k+\Delta'_k+m+p) \Gamma \left(\frac{ \Delta_1+\Delta_2+\Delta_3+\Delta_4-2}{2}\right) \Gamma \left(\frac{\Delta_1+\Delta_4-\Delta_k-\Delta'_k-2 m}{2}\right) \Gamma \left(\frac{\Delta_2+\Delta_3-\Delta_k-\Delta'_k-2m}{2}\right) }. \nonumber 
\end{align}}
It can be explicitly resummed to obtain,

{\scriptsize
\begin{align}
R_m = & -\frac{  \Gamma \left(\frac{\Delta_1+\Delta_4+\Delta_k+\Delta'_k-2}{2}\right)  \Gamma \left(\frac{\Delta_2+\Delta_3+\Delta_k+\Delta'_k-2}{2}\right)\Gamma\left( 1 +m + \frac{\Delta_k + \Delta'_k-\Delta_1 -\Delta_4}{2}\right)\Gamma\left( 1 +m + \frac{\Delta_k + \Delta'_k-\Delta_2 -\Delta_3}{2}\right) }{4 \pi m! \Gamma \left(\frac{1}{2} (\Delta_1+\Delta_2+\Delta_3+\Delta_4-2)\right)}  \nonumber \\
& \pFq{4}{3}{1\quad,-m \quad,\frac{1}{2} (\Delta_2+\Delta_3+\Delta_k+\Delta'_k-2)\quad,\frac{1}{2} (\Delta_1+\Delta_4+\Delta_k+\Delta'_k-2)}{\frac{1}{2} (-\Delta_2-\Delta_3+\Delta_k+\Delta'_k+2)\quad,\frac{1}{2} (-\Delta_1-\Delta_4+\Delta_k+\Delta'_k+2)\quad,m+\Delta_k+\Delta'_k}{-1}.
\end{align}}
For the special case of $\Delta_i = \Delta_k = \Delta'_k=2$ it reduces to,
\begin{align}
    R_m = -\frac{9(3m+4)4^m (m+1)!^2}{2(m+1)(2m+3)!} \alpha^2,
\end{align}
where $\alpha^2 = 1/576 \pi^4$. This is the same Mellin amplitude (for $\lambda=1,R=1$) that was used in \cite{Aharony:2016dwx}. We use the Mellin amplitude (\ref{loopmel}) in the inversion formula (\ref{Mellin2d}) and evaluate the contour integrals at,
\begin{align}
    s=\Delta -J, \qquad t= \Delta_k + \Delta_k^\prime + 2 m.
\end{align}
The residue gives the following mess for the OPE coefficients at order $\frac{1}{N^4}$,

{\tiny
\begin{align}
   a^{(2)}_{0,J} = & \sum_{m=0}^{\infty} \frac{  \Gamma (J+\underline{\Delta}_{12}-1) \Gamma \left(\frac{ \underline{\Delta}_{34} - \underline{\Delta}_{12}}{2}\right) \Gamma \left(\frac{ \underline{\Delta}_{23}+\Delta_k+\Delta'_k-2}{2}\right) \Gamma \left(\frac{ \underline{\Delta}_{14}+\Delta_k+\Delta'_k-2}{2}\right)\Gamma (2 J+\Delta_1+\Delta_2)\Gamma (J+\Delta_1) \Gamma (J+\Delta_2)  \kappa_{\underline{\Delta}_{12} + 2J}}{32 \pi^2 m!  \Gamma (\Delta_1) \Gamma (\Delta_2) \Gamma (\Delta_3) \Gamma (\Delta_4)  } \nonumber \\
   &   \frac{\Gamma \left(\frac{2 m+\Delta_{23}+\Delta_k+\Delta'_k}{2}\right) \Gamma \left(\frac{2 m-\underline{\Delta}_{14}+\Delta_k+\Delta'_k+2}{2}\right) \Gamma \left(\frac{2 m+\Delta_{14}+\Delta_k+\Delta'_k}{2}\right)   \Gamma \left(\frac{2 J+\Delta_1+\Delta_2-\Delta_3+\Delta_4}{2}\right) \Gamma \left(\frac{2 m-\underline{\Delta}_{23}+\Delta_k+\Delta'_k+2}{2}\right)}{ \Gamma \left(\frac{2 J+\Delta_1+\Delta_2}{2}\right)^4} \nonumber \\
   &  \, \pFq{3}{2}{\frac{\underline{\Delta}_{12}-\Delta_{34}+2 J}{2},\frac{\Delta_k+\Delta'_k-\underline{\Delta}_{23}+2 m+2}{2},\frac{\Delta_k+\Delta'_k+\Delta_{23}+2 m}{2}}{\frac{ \Delta_k+\Delta'_k+\Delta_{23}+2 J+2 m+2}{2},\frac{2 \Delta_1+\Delta_{23}+\Delta_k+\Delta'_k+2 J+2 m}{2}}{1} \pFq{4}{3}{1,-m,\frac{\underline{\Delta}_{23}+\Delta_k+\Delta'_k-2}{2},\frac{ \underline{\Delta}_{14}+\Delta_k+\Delta'_k-2}{2}}{\frac{ \Delta_k+\Delta'_k-\underline{\Delta}_{23}+2}{2},\frac{ \Delta_k+\Delta'_k-\underline{\Delta}_{14}+2}{2},\Delta_k+\Delta'_k+m}{-1} . \label{ope44}
\end{align}}

\noindent Here we have used the notation $\underline{\Delta}_{ij} \equiv \Delta_i + \Delta_j$ and  $ \Delta_{ij} \equiv \Delta_i - \Delta_j$. Like before, the leading order contribution in $J$ is obtained from the leading
pole at $m=0$. At large $J$, the OPE coefficient (\ref{ope44}) becomes,

{\tiny
\begin{align}
   a^{(2)}_{0,J\gg 1} = \frac{2^{\underline{\Delta}_{12}+2 J} \Gamma \left(\frac{\underline{\Delta}_{34} -\underline{\Delta}_{12} }{2}\right) \Gamma \left(\frac{ \Delta_{14}+\Delta_k+\Delta'_k}{2}\right) \Gamma \left(\frac{\underline{\Delta}_{14}+\Delta_k+\Delta'_k-2}{2}\right) \Gamma \left(\frac{\Delta_{23}+\Delta_k+\Delta'_k}{2}\right) \Gamma \left(\frac{\underline{\Delta}_{23}+\Delta_k+\Delta'_k-2}{2}\right) }{64 \pi ^{5/2} \Gamma (\Delta_1) \Gamma (\Delta_2) \Gamma (\Delta_3) \Gamma (\Delta_4) \Gamma (\Delta_k+\Delta'_k)J^{\frac{2 \Delta_k+2 \Delta'_k-\sum_i \Delta_i}{2}}} \left( \frac{1}{J^{3/2}} + O \left( \frac{1}{J^{5/2}} \right)  \right).
\end{align}}

\subsubsection{Anomalous dimensions \texorpdfstring{$\frac{1}{N^4}\gamma^{(2)}_{0,J}$}{N}}
For identical external scalars the order $\frac{1}{N^4}$ part of (\ref{cstruc}) is,

{\footnotesize
\begin{align}
    c^t(\Delta,J) \supset \frac{1}{N^4} \left(\frac{a^{(0)}_{n,J} \gamma^{(2)}_{n,J}+a^{(1)}_{n,J}\gamma^{(1)}_{n,J} }{(\Delta -2 \Delta_\phi-J-2n)^2}+\frac{a^{(0)}_{n,J} \left(\gamma^{(1)}_{n,J}\right)^2}{(\Delta -2 \Delta_\phi-J-2n)^3}+\frac{a^{(2)}_{n,J}}{\Delta -2 \Delta_\phi-J-2n}\right).
\end{align} }
\noindent Using the Mellin amplitude (\ref{loopmel}) for identical external scalars in the Mellin inversion formula (\ref{Mellin2d}), we find a double pole corresponding to the double-trace primaries. The reason we do not find a triple pole is because the quartic theory has a vanishing $\gamma^{(1)}_{0,J}$ -- there are no tree level diagrams in the quartic theory! The OPE function simplifies to,
\begin{align}
    c^t(\Delta,J) \supset \frac{1}{N^4} \left(\frac{a^{(0)}_{n,J} \gamma^{(2)}_{n,J} }{(\Delta -2 \Delta_\phi-J-2n)^2}+\frac{a^{(2)}_{n,J}}{\Delta -2 \Delta_\phi-J-2n}\right).
\end{align}
Evaluating the contour integrals we obtain,
{\scriptsize
\begin{align}
    \gamma^{(2)}_{0,J} = & -\sum_{m=0}^{\infty}\, \pFq{3}{2}{\Delta_\phi+J,2 \Delta_\phi+J-1,\Delta_\phi+J}{\frac{\Delta_k+\Delta'_k}{2}+\Delta_\phi+J+m,2 (\Delta_\phi+J)}{1}  \pFq{4}{3}{1,-m,\frac{ \Delta_k+\Delta'_k-2+2\Delta_\phi}{2},\frac{ \Delta_k+\Delta'_k-2+2\Delta_\phi}{2}}{\frac{\Delta_k+\Delta'_k-2 \Delta_\phi+2}{2},\frac{\Delta_k+\Delta'_k-2 \Delta_\phi+2}{2},\Delta_k+\Delta'_k+m}{-1} \nonumber \\
    &\frac{J! \Gamma (J+\Delta_\phi)^2 \Gamma \left(\frac{ \Delta_k+\Delta'_k-2+2\Delta_\phi}{2}\right)^2 \Gamma \left(\frac{2 m+\Delta_k+\Delta'_k}{2}\right)^2 \Gamma \left(\frac{2m+\Delta_k+\Delta'_k-2 \Delta_\phi+2}{2}\right)}{32 \pi ^4 m! \Gamma (\Delta_\phi)^2} . \label{gamm4}
\end{align}}
Unlike the tree diagrams of the cubic theory, the satellite poles here do not truncate for any value of $\Delta_k$ or $\Delta_k^\prime$. For the special case of $\Delta_\phi=\Delta_k=\Delta'_k$ we have,
{\small
\begin{align}
    \gamma^{(2)}_{0,J} = & - \frac{4^{\Delta_\phi-4} \Gamma \left(\frac{2\Delta_\phi-1}{2}\right)^2 \Gamma (J+1) \Gamma (J+\Delta_\phi)^2}{\pi^5} \sum_{m=0}^\infty \Gamma (m+\Delta_\phi)^2 \times \nonumber \\
    &  \pFq{3}{2}{m+1 \, ,\frac{2\Delta_\phi-1}{2} \, ,\Delta_\phi \, }{1 \, \, ,m+2\Delta_\phi \, }{1} \, \pFq{3}{2}{J+\Delta_\phi \, ,J+2\Delta_\phi-1 \, ,J+\Delta_\phi \, }{J+m+2\Delta_\phi \, ,2 (J+\Delta_\phi)\,}{1}  \label{gamm4i}
\end{align}}
Like before, the $u$-channel contribution cancels the $t$-channel contribution for odd spins and adds up for even. We find that our results match numerically with those of Aharony et al. \cite{Aharony:2016dwx}. They calculated the contribution of the bubble diagram to the anomalous dimension of double-trace primaries in the $\phi^4$ theory for $\Delta_\phi=2$ in $d=2,4$ and $J=2,4$. Our result, even when we sum the first few satellite poles matches their result with less than one percent error. 

The large $J$ limit of our result is tractable for the same reasons mentioned near (\ref{largeJ}). Using (\ref{3f2asymp}) and adding the $u$-channel result we get for $\Delta_\phi=2$,
\begin{align}
  \gamma^{(2)}_{0,J} = -\frac{\left(1+(-1)^J\right)3 \alpha^2}{J^4} \left( 1 -\frac{6}{J} +\frac{119}{5 J^2} -\frac{396}{5 J^3} +\frac{8399}{35 J^4} -\frac{24054}{35 J^5} + \frac{66413}{35 J^6} \ldots \right).
\end{align}
This matches exactly with \cite{Aharony:2016dwx}, when written in terms of the conformal spin $\tilde J$.

\subsection{Results for \texorpdfstring{$d=4$}{d4}}
The results for $d=4$ can be derived similarly. Like in the previous sections, we only state the $t$-channel contribution since the $u$-channel contribution can be simply obtained by $\Delta_1 \leftrightarrow \Delta_2$.
\\

\vspace{0.2cm} 
\noindent \underline{\textbf{Contact diagram}} \\
As before we find,
\begin{align*}
    c_{contact}= 0,
\end{align*}
for the same reason mentioned near (\ref{contact}).
\\

\vspace{0.2cm} 
\noindent \underline{\textbf{Exchange diagram}} \\
The OPE coefficient for the double-trace primary is given by,

{\tiny
\begin{align}
    a^{(1)}_{0,J}= & \sum_{m=0}^{\infty}  \frac{-\pi^2 \kappa_{\Delta_1 + \Delta_2 + 2 J}\Gamma (\Delta_k-1) \Gamma (\Delta_k) \Gamma (J+\Delta_1+\Delta_2-2) \Gamma (2 J+\Delta_1+\Delta_2) \Gamma \left(\frac{\Delta_{32}+\Delta_{41}}{2}\right) }{m! \Gamma (m+\Delta_k-1) \Gamma \left(\frac{ \Delta_{14}+\Delta_k}{2}\right) \Gamma \left(\frac{\Delta_{41}+\Delta_k}{2}\right) \Gamma \left(\frac{\Delta_{23}+\Delta_k}{2}\right) \Gamma \left(\frac{ \Delta_{32}+\Delta_k}{2}\right)  \Gamma \left(\frac{2 J+\Delta_1+\Delta_2+\Delta_3-\Delta_4}{2}\right)  } \nonumber \\
    & \times \Bigg \{ \, \frac{(\Delta_k-\Delta_1-\Delta_4+2 m+2) (\Delta_k-\Delta_2-\Delta_3+2 m+2)}{2} \, \pFq{3}{2}{\frac{2 J+\Delta_1+\Delta_2-\Delta_3+\Delta_4}{2},\frac{2 m-\Delta_2-\Delta_3+\Delta_k+4}{2},\frac{2 m+\Delta_2-\Delta_3+\Delta_k}{2}}{\frac{2 J+2 m+\Delta_2-\Delta_3+\Delta_k+4}{2},\frac{2 J+2 m+2 \Delta_1+\Delta_2-\Delta_3+\Delta_k}{2}}{1} \nonumber \\
    & -2 \, \pFq{3}{2}{\frac{2 J+\Delta_1+\Delta_2-\Delta_3+\Delta_4}{2},\frac{2 m-\Delta_2-\Delta_3+\Delta_k+2}{2},\frac{2 m+\Delta_2-\Delta_3+\Delta_k-2}{2}}{\frac{2 J+2 m+\Delta_2-\Delta_3+\Delta_k+2}{2},\frac{2 J+2 m+2 \Delta_1+\Delta_2-\Delta_3+\Delta_k-2}{2}}{1} \Bigg \} \frac{\Gamma \left(\frac{2 m+\Delta_{14}+\Delta_k}{2}\right) \Gamma \left(\frac{2 m+\Delta_{23}+\Delta_k}{2}\right)}{\Gamma \left(\frac{\Delta_1+\Delta_4-\Delta_k-2m}{2}\right) \Gamma \left(\frac{\Delta_2+\Delta_3-\Delta_k-2m}{2}\right)}.
\end{align}}
\noindent The anomalous dimension is given by,
{\tiny
\begin{align}
 \gamma^{(1)}_{0,J} = & - \sum_{m=0}^{\infty} \frac{ \kappa_{2\Delta_\phi +2J}  J! \Gamma \left(\frac{\Delta_k-1}{2}\right) \Gamma \left(\frac{\Delta_k+1}{2}\right) \Gamma (\Delta_\phi)^2 2^{2 \Delta_k+4 \Delta_\phi+4 J} \Gamma \left(J+\Delta_\phi-\frac{1}{2}\right) \Gamma \left(J+\Delta_\phi+\frac{1}{2}\right) \Gamma \left(m+\frac{\Delta_k}{2}\right)^2 }{ 16 m! \Gamma \left(\frac{\Delta_k}{2}\right)^2 (2 \Delta_\phi+J-2) \Gamma (J+\Delta_\phi) \Gamma (m+\Delta_k-1) \Gamma \left(-m+\Delta_\phi-\frac{\Delta_k}{2}\right)^2} \nonumber \\
 & \left(\, \pFq{3}{2}{m+\frac{\Delta_k}{2}-1,m+\frac{\Delta_k}{2}-\Delta_\phi+1,J+\Delta_\phi}{J+m+\frac{\Delta_k}{2}+1,J+m+\frac{\Delta_k}{2}+\Delta_\phi-1}{1}-\frac{ (\Delta_k-2 \Delta_\phi+2 m+2)^2}{4} \, \pFq{3}{2}{m+\frac{\Delta_k}{2},m+\frac{\Delta_k}{2}-\Delta_\phi+2,J+\Delta_\phi}{J+m+\frac{\Delta_k}{2}+2,J+m+\frac{\Delta_k}{2}+\Delta_\phi}{1}\right). \label{gamma4d}
\end{align}}

\noindent We can obtain closed form expressions when the satellite poles truncate. For e.g. when $\Delta_k=2\Delta_\phi - 2$, 
\begin{align}
    \gamma^{(1)}_{0,J}= -\frac{(\Delta_\phi-1) \Gamma (2 \Delta_\phi-1) \Gamma (J+1)}{\Gamma (2 \Delta_\phi+J-1)}.
\end{align}
All of our results match with \cite{Liu:2018jhs}. 
\\

\vspace{0.2cm} 
\noindent \underline{\textbf{Bubble diagram}} \\
The anomalous dimension of the double-trace primary for the $\phi^4$ theory at order $\frac{1}{N^4}$ is given by,

{\tiny
\begin{align}
    \gamma^{(2)}_{0,J} = & \sum_{m=0}^{\infty} \sum_{p=0}^{m}\frac{-(p+1) (\Delta_\phi+p-1)^2 (2 \Delta_\phi+p-3) \Gamma (\Delta_\phi)^2  \Gamma (J+\Delta_\phi)^2 \Gamma (m+\Delta_\phi)^2 \Gamma (p+2 \Delta_\phi-2)^2 J!  }{32 \pi ^8 (2 \Delta_\phi+2 p-3) (2 \Delta_\phi+2 p-1)  (2 \Delta_\phi+J-2) \Gamma (\Delta_\phi-1)^4  \Gamma (p+1)^2 \Gamma (m-p+1) \Gamma (m+p+2 \Delta_\phi-1)} \nonumber \\  
    & \left(\Gamma (m+1) \,  \pFq{3}{2}{J+\Delta_\phi,J+2 \Delta_\phi-2,J+\Delta_\phi}{J+m+2 \Delta_\phi-1,2 (J+\Delta_\phi)}{1}-\Gamma (m+2) \, \pFq{3}{2}{J+\Delta_\phi,J+2 \Delta_\phi-2,J+\Delta_\phi}{J+m+2 \Delta_\phi,2 (J+\Delta_\phi)}{1} \right) ,
\end{align}}

\noindent where the sum over $p$ can be performed explicitly using the ${}_4F_3$ hypergeometric function. Since that is equally uninspiring we leave the sum as it is, and check that our results match numerically with \cite{Aharony:2016dwx}. At large $J$, adding the $u$-channel contribution and using (\ref{3f2asymp}) we get for $\Delta_\phi=2$,
\begin{align}
     \gamma^{(2)}_{0,J} = -\frac{\left(1+(-1)^J\right)6 \alpha ^2}{J^4} \left(1 -\frac 6 J +\frac{133}{5 J^2}-\frac{522}{5 J^3}+\frac{13563}{35 J^4}-\frac{48978}{35 J^5} + \frac{24917}{5 J^6} \ldots \right).
\end{align}
This is identical to \cite{Aharony:2016dwx} when written in terms of the conformal spin.

\section{The flat space limit of the inversion formula and future directions \label{sec4}}
In this note we provided a new perspective to the inversion formula using Mellin space. We showed that in this formulation it is relatively easy to get nontrivial, albeit cumbersome results about anomalous dimensions and OPE coefficients in large $N$ strongly coupled CFTs. In this work we focused only on scalar exchanges and a limited set of Witten diagrams. It would be interesting to include spinning operators and more complicated diagrams. The Mellin amplitude of several diagrams have been recently calculated (see \cite{Yuan:2018qva} and references therein), and can be used in our formula to obtain CFT data beyond $1/N^4$. We also only worked with the leading-twist primary $n=0$, but it would be useful to generalize our results to subleading twists $n>0$.

As mentioned in the introduction, one of our major motivations to formulate the inversion formula in Mellin space was the simplicity of Mellin amplitudes in large $N$ CFTs. However, there is another feature that might make the Mellin space formalism more useful. It was shown by Penedones \cite{Penedones:2010ue} that in the limit of large $s,t$ and massless external particles the Mellin amplitude of a large $N$ CFT is related to the flat space amplitude by\footnote{Recently, a similar result concerning massive external particles was found by Paulos et al. \cite{Paulos:2016fap}. In fact, the corresponding relation to (\ref{con11}) for massive particles is simpler, and it might be easier to connect our results with the massive Froissart-Gribov formula.},
\begin{align}
A(s_{ij}) & = \lim_{R \rightarrow \infty} \frac{R^{3-d}}{\Gamma \left ( \sum_i \Delta_i - \frac d 2 \right) } \int_0^\infty d\beta \beta^{\frac 1 2 \sum_i \Delta_i - \frac d  2 - 1} e^ {-\beta} \mathcal T \left( S_{ij}=\frac{2 \beta}{R^2} s_{ij} \right). \label{con11}  
\end{align}
Where $\mathcal T(S,T)$ is the connected flat space amplitude in ordinary QFT and can be decomposed via its partial waves,
\begin{align}
\mathcal T (S,T) = \sum_J a_J (S)  P_J(\theta). \label{qft}
\end{align}
The Froissart-Gribov (FG) formula,
\begin{align}
a_J(S) = \int_1^\infty d(\cosh \eta) (\sinh \eta)^{d-4} Q_J (\cosh \eta) \text{Disc}_T \mathcal T(S,T), \quad T=\frac{S}{2} (\cosh \eta - 1), \label{FG}
\end{align}
provides the decomposition coefficients $a_J(S)$ in terms of the discontinuity of the flat space scattering amplitude. A natural question to ask is whether the Mellin inversion formula (\ref{Mellin4d}) reduces to the flat space $d=4$ FG formula (\ref{FG}) in the flat space limit (\ref{con11}). A similar question also exists in the bulk-point limit of the position space inversion formula \cite{Maldacena:2015iua,Gary:2009ae,Alday:2017vkk}. However, we believe that it might be easier to approach this problem in Mellin space, due to the apparent simplicity of the flat-space limit and AdS correlators in Mellin space.

Costa et al. \cite{Costa:2012cb} have already established the relationship between the flat space decomposition coefficients $a_J(S)$ and the OPE coefficients $b_J(\nu^2)$, 
\begin{align}
a_J(S) & = \lim_{R\rightarrow \infty} \frac{R^{d-3}}{\mathcal N_1}  \left( \frac{R^2 S}{4} \right)^J \braket{b_J}_S, \qquad \braket{b_J}_S  = \int dx \delta_L(x) b_J(-R^2 S + x). \label{con3}
\end{align}
$S$ and $T$ are the Mandelstam invariants of the flat space QFT, and $\mathcal N_1$ is a normalisation coefficient \cite{Costa:2012cb}. The $b_J(\nu^2)$ are the decomposition coefficients of the conformal four-point function,
\begin{align}
\mathcal{G} (u,v) & = \sum^\infty_{J=0} \int_{-\infty}^{\infty} \! \! \! d\nu \, b_J(\nu^2) F_{\nu,J} (u,v), \label{cpw} 
\end{align}
where $F_{\nu,J}(u,v)$ are the conformal partial waves. The $b_J(\nu^2)$ are simply related to the $c(\Delta,J)$ of the inversion formula by,
\begin{align}
b_J\left( - \left(\Delta -   d/ 2\right)^2 \right)  = \frac{K_{\Delta,J}}{4\pi \left(\Delta - d / 2 \right)}  c(\Delta,J), \label{dic}
\end{align}
where,
{\scriptsize
\begin{align}
    K_{\Delta,J} = \frac{\Gamma(\Delta+1-d/2)}{4^{J-1} \pi \Gamma\left(\Delta-1\right)\kappa_{\Delta+J}\Gamma\left( \frac{\Delta_1 + \Delta_2 - \Delta+J}{2} \right)\Gamma\left( \frac{\Delta_3 + \Delta_4 - \Delta+J}{2} \right)\Gamma\left( \frac{\Delta_1 + \Delta_2 + \Delta+J-d}{2} \right)\Gamma\left( \frac{\Delta_3 + \Delta_4 + \Delta+J-d}{2} \right)}.
\end{align}}

\noindent Thus the question becomes the following -- does the Mellin inversion formula (\ref{Mellin4d}) for $b_J$ in (\ref{dic}), reduce to the the FG formula (\ref{FG}) in the flat space limit (\ref{con3})? We hope to pursue this direction in future work.

\acknowledgments

It is a pleasure to thank Raghu Mahajan and Xinan Zhou for numerous discussions, help and encouragement without which this project would have been impossible. I would like to thank Eric Perlmutter and Massimo Taronna for discussions and important comments on the draft. I would also like to thank Richard Nally for his help in drawing Witten diagrams, and his insufferable sense of humor.  This work was partly supported by the National Science Foundation grant PHY-1720397. Finally, I would like to thank my previous PhD advisor -- Joe Polchinski, an exceptional scientist and a great human being. This paper contains far too many hypergeometric functions than he would have approved of, but may this be my humble tribute to him.


\appendix
\renewcommand{\thesection}{\Alph{section}}

\section{Normalisation of Witten diagrams}\label{normwitt}
The formulas in the main text (\ref{ope2d}), (\ref{gamma}) and (\ref{gamma4d}) do not seem to depend on the bulk cubic coupling constant $g$ or the AdS radius $R$. This is due to our normalization convention that we now elaborate. The four-point function of scalars in any CFT can be written as,
\begin{align}
\braket{\phi_1 \phi_2 \phi_3 \phi_4} & = \frac{1}{(z \zb)^{\frac{\Delta_1 + \Delta_2}{2}}} \sum_{\Delta_p,l_p} f_{\phi \phi O_p}^2 g_{\Delta_p, l_p} (z,\zb) \label{schan} \\
& = \frac{1}{((1-z)(1- \zb))^{\frac{\Delta_2 + \Delta_3}{2}} } \sum_{\Delta_k,l_k} f_{\phi \phi O_k}^2 g_{\Delta_k, l_k} (1-z,1-\zb) \label{tchan} \\
& =   W^t +  W^s + W^u \label{Witten},
\end{align}
where $W^t$ is the sum over all $t$-Witten diagrams, and similarly for $s$ and $u$. We will work with the $t$-channel expansion (\ref{tchan}) in the limit $v \rightarrow 0$ i.e. $\zb \rightarrow 1$. In this limit, the $t$-channel conformal blocks can be expanded in powers of $(1-\zb)$,

{\scriptsize
\begin{align}
     g_{\Delta_k, J_k} \! (1-z,1-\zb) = \! (1-z)^{\frac{\Delta_k+J_k}{2}} \! (1-\zb)^{\frac{\Delta_k-J_k}{2}}  \! \! \left(\! {}_2F_1\left(\frac{ \Delta_{14} + J_k+\Delta_k}{2},\frac{ \Delta_{23} + J_k+\Delta_k }{2};J_k+\Delta_k;1-z \! \right) + O(1-\zb)\! \right)\! . \label{gser}
\end{align}}
The leading term in $t$-channel expansion corresponding to the exchange of a single trace operator $O_k$ is thus given by
{
\begin{align}
   f_{\phi \phi O_k}^2 \frac{(1-z)^{\frac{\Delta_k+J_k}{2}} (1-\zb)^{\frac{\Delta_k-J_k}{2}}}{((1-z)(1- \zb))^{\frac{\Delta_2 + \Delta_3}{2}} }   \, _2F_1\left(\frac{ \Delta_{14} + J_k+\Delta_k}{2},\frac{ \Delta_{23} + J_k+\Delta_k }{2};J_k+\Delta_k;1-z\right).  \label{tser}
\end{align}}
For a large $N$ theory, the four-point function can also be calculated perturbatively in $1/N$. At leading order, the four-point function in the cubic theory is given by the sum over all tree level exchange Witten diagrams $W_{\Delta_k,J_k}$ in (\ref{Witten}) as shown in figure \ref{stu}. As explained in the main text, the tree level exchange diagrams carry information about the double-trace exchanges in addition to the single-trace exchanges. In particular, the $s$ and $u$-Witten diagrams when expanded in the $t$-channel only consist of double-trace exchanges \cite{Heemskerk:2009pn}. Thus in terms of the $t$-channel expansion, (\ref{Witten}) comprises of a single-trace $t$-channel exchanges and several multi-trace exchanges.

We choose the following convention for the normalisation of $g$ -- the contribution of the single-trace exchange $O_k$ in the $t$-Witten diagram in the limit $\zb \rightarrow 1$ should exactly equal (\ref{tser}) with the OPE coefficient $f^2_{\phi \phi O_k}$ set to one. We begin with the Mellin representation of the exchange diagram,

{\scriptsize
\begin{align}
W^t_{\Delta_k,J_k}(u,v)  & = - \frac{1}{u^{\frac{\Delta_1 + \Delta_2}{2}}} \int \frac{ds dt}{(2\pi i )^2} u^{\frac s 2} v^{\frac{t - \Delta_2 - \Delta_3}{2}}  \Gamma \left(\frac{\Delta_1 + \Delta_4 -t}{2} \right) \Gamma \left(\frac{\Delta_2 + \Delta_3 -t}{2} \right) \Gamma \left( \frac{\Delta_1 + \Delta_2 -s}{2}\right) \nonumber  \\
& \times \Gamma\left(\frac{ \Delta_3 + \Delta_4 - s}{2} \right) \Gamma\left( \frac{s+t-\Delta_2 - \Delta_4 }{2} \right) \Gamma\left(\frac{s+t-\Delta_1 - \Delta_3 }{2}\right) \mathcal N g^2 R^{5-d}\sum_{m=0} \frac{R_m}{t - \Delta_k - 2m}.  \nonumber \\
R_m & =  \frac{\Gamma\left(  \frac{\Delta_1 + \Delta_4 + \Delta_k - d}{2} \right)\Gamma\left(  \frac{\Delta_2 + \Delta_3 + \Delta_k - d}{2} \right) }{2\Gamma\left( \frac{\Delta_1+\Delta_2+\Delta_3+\Delta_4 -d}{2} \right)} \frac{\left(\frac{1}{2} (\Delta -\Delta_1-\Delta_4)+1\right)_m \left(\frac{1}{2} (\Delta -\Delta_2-\Delta_3)+1\right)_m}{m! \Gamma \left(m+\Delta +1 - \frac d 2\right)} \label{twitt}
\end{align}}

\noindent In the limit $\zb \rightarrow 1$ i.e. $v \rightarrow 0$, $m\neq0$ terms contribute at subleading orders in $v$ and thus we can set $m=0$. Closing the $t$ integration contour to the right and only picking up the pole at $t=\Delta_k$ we have,

{\scriptsize
\begin{align}
\left[W^t(u,v) \right] & = - \frac{1}{u^{\frac{\Delta_1 + \Delta_2}{2}}} \int \frac{ds}{(2\pi i )} u^{\frac s 2} v^{\frac{\Delta_k - \Delta_2 - \Delta_3}{2}}  \Gamma \left(\frac{\Delta_1 + \Delta_4 - \Delta_k}{2} \right) \Gamma \left(\frac{\Delta_2 + \Delta_3 -  \Delta_k}{2} \right) \Gamma \left( \frac{\Delta_1 + \Delta_2 -s}{2}\right) \nonumber  \\
& \times \Gamma\left(\frac{ \Delta_3 + \Delta_4 - s}{2} \right) \Gamma\left( \frac{s+ \Delta_k-\Delta_2 - \Delta_4 }{2} \right) \Gamma\left(\frac{s+ \Delta_k-\Delta_1 - \Delta_3 }{2}\right) \mathcal N g^2 R^{5-d} R_0. \nonumber
\end{align}}
The square brackets denote the fact that we have ignored all the double-trace poles when closing the contour. Making a coordinate change $s \rightarrow -2s + \Delta_1 + \Delta_2$ and using the Mellin-Barnes representation of the hypergeometric function ${}_2 F_1$ the above becomes 

{\footnotesize
\begin{align}
W^t(u,v) & = 2  v^{\frac{\Delta_k - \Delta_2 - \Delta_3}{2}}  \mathcal N g^2 R^{5-d} R_0 
 \Gamma \left(\frac{\Delta_1 + \Delta_4 - \Delta_k}{2} \right) \Gamma \left(\frac{\Delta_2 + \Delta_3 -  \Delta_k}{2} \right)  \Gamma\left(  \frac{\Delta_k}{2} + \frac{\Delta_{14}}{2} \right) \nonumber  \\
& {}_2 F_1 \left(  \frac{\Delta_k+\Delta_{14}}{2}, \frac{\Delta_k+\Delta_{23}}{2}, \Delta_k,1-u  \right) \frac{ \Gamma \left(  \frac{\Delta_k+\Delta_{23}}{2} \right) \Gamma\left( \frac{\Delta_k - \Delta_1 + \Delta_4 }{2} \right) \Gamma\left( \frac{\Delta_k - \Delta_2 + \Delta_3 }{2} \right)}{\Gamma\left(\Delta_k\right)} \nonumber
 \end{align}}
Rewriting this in terms of $z,\zb$ and taking the limit $\zb \rightarrow 1$ and comparing it with (\ref{tser}) for scalar exchange $J_k =0$ we find,
\begin{align}
g^{-2} = & 2 R^{5-d}  \mathcal NR_0 
 \Gamma \left(\frac{\Delta_1 + \Delta_4 - \Delta_k}{2} \right) \Gamma \left(\frac{\Delta_2 + \Delta_3 -  \Delta_k}{2} \right)  \Gamma\left(  \frac{\Delta_k}{2} + \frac{\Delta_1 - \Delta_4}{2} \right) \nonumber  \\
 & \frac{ \Gamma \left(  \frac{\Delta_k}{2} + \frac{\Delta_2 - \Delta_3}{2} \right) \Gamma\left( \frac{\Delta_k - \Delta_1 + \Delta_4 }{2} \right) \Gamma\left( \frac{\Delta_k - \Delta_2 + \Delta_3 }{2} \right)}{\Gamma\left(\Delta_k\right)}, \label{wittnorm}
\end{align}
where $R_0$ is given in (\ref{twitt}) and $\mathcal N$ was defined in (\ref{norm}) as,
\begin{align}
\mathcal N = \frac{\pi^{d/2}}{2}\Gamma \left(  \frac{\sum_{i=1}^n \Delta_i - \frac d 2}{2} \right) \prod_{i=1}^n \frac{1}{2\pi^{d/2} \Gamma(\Delta_i- \frac d 2+1) } .
\end{align}
The normalization \ref{wittnorm} is particularly convenient, since it allows us to directly compare our results with those of Liu et al. \cite{Liu:2018jhs}.

\section{Anomalous dimensions via collinear blocks} \label{coll}
In this appendix, we obtain the anomalous dimensions of the double-trace primaries using the collinear block decomposition of the four-point function. Our aim is to show the equivalence of our results from the Mellin inversion formula (\ref{gamma}) with the results obtained in this appendix using standard techniques. 

In the light cone limit $u \ll v \ll 1$ the four-point function can be decomposed in terms of the collinear conformal blocks
{\small
\begin{align}
    \braket{\phi \phi \phi \phi} = \frac{1}{u^{\Delta_\phi}} \left(  u^{\frac{\tau_{min}}{2}} \sum_{J} a_{\Delta,J}(1-v)^J {}_2F_1(\Delta+J,\Delta+J,2\Delta+2J,1-v) + \ldots \right). \label{collexp}
\end{align}}
where the $\ldots$ represent higher orders in $u$. For large $N$ theories of our interest, the minimal twist operators are precisely the double-trace primaries that have the scaling dimension,
\begin{align}
    \Delta=2\Delta_\phi + J + 2n + \gamma,
\end{align}
where $\gamma$ is the anomalous dimension. Like in the main text, the CFT data can be expanded perturbatively in $\frac{1}{N}$ as,
\begin{align}
    a_{n,J} = a^{(0)}_{n,J} + \frac{a^{(1)}_{n,J}}{N^2}  + \frac{a^{(2)}_{n,J}}{N^4}  + \ldots \, , \qquad \gamma_{n,J} =  \frac{\gamma^{(1)}_{n,J}}{N^2}  + \frac{\gamma^{(2)}_{n,J}}{N^4}  + \ldots. \, .
\end{align}
At order $\frac{1}{N^2}$ the $\log u$ part of (\ref{collexp}) is given by,
{\footnotesize
\begin{align}
    \braket{\phi \phi \phi \phi } \supset \frac{\log u}{N^2} \sum_J a^{(0)}_{n,J} \gamma^{(1)}_{n,J}  (1-v)^J \, _2F_1(2 \Delta_\phi+2 J,2 \Delta_\phi+2 J;4 \Delta_\phi+4 J;1-v). \label{collexp1}
\end{align}}

\noindent The idea is to calculate the left hand side using Witten diagrams and compare with the right hand side to read off the anomalous dimensions. In the bulk scalar theory with cubic interactions, the order $\frac{1}{N^2}$ contribution comes from the exchange diagram, which we now calculate in Mellin space.

{\footnotesize
\begin{align}
    \braket{\phi\phi\phi\phi} \supset \frac{1}{u^{\Delta_\phi}} \int \frac{ds dt}{(2\pi i )^2} u^{s/2} v^{\frac{t}{2}-\Delta_\phi} \Gamma \left(\frac{2 \Delta_\phi-s}{2}\right)^2 \Gamma \left(\frac{2 \Delta_\phi-t}{2}\right)^2  \Gamma \left(\frac{s+t-2 \Delta_\phi}{2}\right)^2 A(s,t) \label{collmel}
\end{align}}
Using the Mellin amplitude from the main text for the tree level $t$ exchange diagram (\ref{melexc}) and the normalization (\ref{wittnorm}) we have,
\begin{align}
    A(s,t) = \sum_{m=0}^{\infty}  \frac{ \Gamma (\Delta_k)^2  \left(\frac{\Delta_k}{2}-\Delta_\phi+1\right)_m^2}{m! 2\Gamma \left(\frac{\Delta_k}{2}\right)^4  \Gamma \left(\Delta_\phi-\frac{\Delta_k}{2}\right)^2 \Gamma (m+\Delta_k) (\Delta_k+2 m-t)}. \label{ast}
\end{align}
The $s$ contour integral can be evaluated at the double pole $s=2\Delta_\phi$, which subsequently gives a term with $ \log u$. The remaining $t$ integral can be evaluated by closing the contour on the left and summing over the infinitely many poles $t=-2p$ due to the $\Gamma\left(\frac{s+t-2\Delta_\phi}{2}\right)^2$ in the Mellin measure. Comparing that with RHS of (\ref{collexp1}) gives,

{\footnotesize
\begin{align}
    & \sum_{J=0}^{\infty} a^{(0)}_{0,J} \gamma^{(1)}_{0,J}  (1-v)^J \, _2F_1(2 \Delta_\phi+2 J,2 \Delta_\phi+2 J;4 \Delta_\phi+4 J;1-v) \nonumber \\
    & =\sum_{m=0}^\infty \sum_{p=0}^\infty  \frac{2 \Gamma (\Delta_k)^2  \Gamma (p+\Delta_\phi)^2   \left(2+(\Delta_k+2 (m+p)) \left(-2 H_{p+\Delta_\phi-1}+2 H_p+\log (v)\right)\right)}{v^{\Delta_\phi+p} m! (p!)^2 \Gamma \left( \Delta_\phi - \frac{\Delta_k}{2} - m \right)^2 \Gamma \left(\frac{\Delta_k}{2}\right)^4  \Gamma (m+\Delta_k) (\Delta_k+2 (m+p))^2} \label{eq}
\end{align}}
where $H_n$ is the harmonic number. We would like to invert this equation to obtain the anomalous dimensions. There are several ways of doing it, one is to use the orthogonality of the hypergeometric function, as is shown in \cite{Aharony:2016dwx}. The other is to use Mathematica to expand both sides in $(1-v)$, and read off the coefficients $a^{(0)}_{0,J} \gamma^{(1)}_{0,J}$ for every term in $(1-v)$. We shall use the latter. The mean field theory coefficients $a^{(0)}_{n,J}$ are given in (\ref{mft}). 

The sum over $m$ in (\ref{eq}) corresponds to the sum over satellite poles that was also found in the main text. We solve (\ref{eq}) for an arbitrary satellite pole $m=k$ (for some positive integer $k$) and find that it matches with the $m=k$ term of (\ref{gamma}). Thus we show that our results using the Mellin inversion formula matches term by term with the results from the standard techniques of collinear block decomposition. 

The satellite poles truncate when $2\Delta_\phi - \Delta_k = 2 \mathbb Z^+$. We take $\Delta_k =3,\Delta_\phi=3$ i.e. no truncation, and $k=2$. The sum over $p$ can be explicitly performed after some massaging, and we find that the right hand side of (\ref{eq}) becomes

{\tiny
\begin{align}
     \frac{225}{48 \pi^3 v^3} \left(\Phi \left(\frac{1}{v},2,\frac{7}{2}\right)+\frac{2 v^3 \log (v) \left(v (607-15 v (5 v (3 v-11)+73))+225 \sqrt{v} (v-1)^4 \coth ^{-1}\left(\sqrt{v}\right)-64\right)-4 (v-1) v (v (17 v-2)+9)}{225(v-1)^4}\right)
\end{align}}
where $\Phi$ is the Lerch Zeta function. Using the following recursion relation,
\begin{align}
    \Phi(z,s,\alpha+1) = \frac{1}{z} \left(\Phi(z,s,\alpha) - \frac{1}{\alpha^s} \right),
\end{align}
and comparing both sides of (\ref{eq}) order by order in  $(1-v)$ we obtain,
\begin{align}
    \gamma^{(1)}_{0,J} \supset - \frac{75}{16 \pi ^2}  \Gamma (J+1) \Gamma (J+3) \, _3\tilde{F}_2\left(\frac{3}{2},\frac{7}{2},J+3;J+\frac{9}{2},J+\frac{13}{2};1\right).
\end{align}
This exactly matches with $m=2$ part of (\ref{gamma}).

\bibliographystyle{JHEP}
\bibliography{biblio}
\end{document}